\documentclass[twocolumn,prb,aps,showpacs,floatfix,superscriptaddress,preprintnumbers,amsmath]{revtex4-1}

\usepackage{amsmath}
\usepackage{graphicx}
\usepackage{SIunits}
\usepackage{xspace}
\usepackage{wasysym}
\usepackage{tabularx}

\usepackage{hyperref}
\hypersetup{
    unicode=false,          
    pdftoolbar=true,        
    pdfmenubar=true,        
    pdffitwindow=false,     
    pdfstartview={FitH},    
    pdftitle={My title},    
    pdfauthor={Author},     
    pdfsubject={Subject},   
    pdfcreator={Creator},   
    pdfproducer={Producer}, 
    pdfkeywords={keyword1} {key2} {key3}, 
    pdfnewwindow=true,      
    colorlinks=true,       
    linkcolor=blue,          
    citecolor=blue,        
    filecolor=blue,      
    urlcolor=blue,       
    plainpages=false        
}

\usepackage{color}

\newcommand{\Tc}{$T_{\rm{C}}$\xspace}
\newcommand{\Ts}{$T_{\rm{s}}$\xspace}
\newcommand{\Th}{$T_{\rm{h}}$\xspace}
\newcommand{\wn}{$\rm{cm}^{-1}$\xspace}
\newcommand{\cps}{$\rm{counts\:(s \cdot \milli\watt)^{-1}}$\xspace}
\newcommand{\vq}{\textbf{q}}

\begin{document}

\renewcommand{\tabularxcolumn}[1]{>{\normalsize\centering\arraybackslash}m{#1}}

\title{A Raman study of the temperature and magnetic field dependence of electronic and lattice properties in MnSi}

\author{H.-M.~Eiter}
\affiliation{Walther Meissner Institute, Bavarian Academy of Sciences and Humanities, 85748 Garching, Germany}

\author{P.~Jaschke}
\affiliation{Walther Meissner Institute, Bavarian Academy of Sciences and Humanities, 85748 Garching, Germany}

\author{R.~Hackl}
\affiliation{Walther Meissner Institute, Bavarian Academy of Sciences and Humanities, 85748 Garching, Germany}

\author{A.~Bauer}
\affiliation{Physics Department E21, Technical University Munich, 85748 Garching, Germany}

\author{M.~Gangl}
\affiliation{Physics Department E21, Technical University Munich, 85748 Garching, Germany}

\author{C.~Pfleiderer}
\affiliation{Physics Department E21, Technical University Munich, 85748 Garching, Germany}

\date{\today}

\begin{abstract}
The temperature and magnetic field dependence of lattice and carrier excitations in MnSi is studied in detail using inelastic light scattering. The pure symmetry components of the electronic response are derived from the polarization dependent spectra. The $E$ and $T_2$ responses agree by and large with longitudinal and optical transport data. However, an anomaly is observed right above the magnetic ordering temperature $T_{\rm C}=29$\,K that is associated with the fluctuations that drive the transition into the helimagnetic phase first order. The $T_1$ spectra, reflecting mostly chiral spin excitations, have a temperature dependence similar to that of the $E$ and $T_2$ symmetries. The response in the fully symmetric $A$ representation has a considerably weaker temperature dependence than that in the other symmetries. All nine Raman active phonon lines can be resolved at low temperature. The positions and line widths of the strongest four lines in $E$ and $T_2$ symmetry are analyzed in the temperature range $4<T<310$\,K. Above 50\,K, the temperature dependence is found to be conventional and given by anharmonic phonon decay and the lattice expansion. Distinct anomalies are observed in the range of the helimagnetic transition and in the ordered phase. Applying a magnetic field of 4\,T, well above the critical field, removes all anomalies and restores a conventional behavior highlighting the relationship between the anomalies and magnetism. The anomaly directly above \Tc in the fluctuation range goes along with an anomaly in the thermal expansion. While the lattice constant changes continuously and has only a kink at \Tc, all optical phonons soften abruptly suggesting a direct microscopic coupling between spin order and optical phonons rather than a reaction to magnetostriction effects.

\end{abstract}


\maketitle

\section{Introduction}

A long-standing problem in the field of itinerant-electron magnetism concerns the interplay of collective excitations such as phonons, plasmons or magnons with the conduction electrons.  While elements such as Fe, Ni and Co are clearly prototypical for itinerant-electron ferromagnetism and Cr is prototypical for spin-density and charge density wave order, their relatively high transition temperatures imply large energy scales for the collective excitations that result in a complex interplay. A materials class that has played a major role in the 1970s and 80s for the development of the present day understanding of itinerant-electron magnetism are weakly magnetic transition metal compounds such as ZrZn$_{2}$ or Ni$_{3}$Al. For their description, a self-consistent phenomenological model was developed taking into account dispersive spin fluctuations associated with the damping due to the particle-hole continuum.\cite{1985:Moriya:Book, 1985:Lonzarich:JPhysCSolidState} This spin-fluctuation theory is in excellent quantitative agreement with experiment and, during the following decade, became the basis for the description of magnetic quantum phase transitions.\cite{2001:Stewart:RevModPhys, 2007:Lohneysen:RevModPhys}

Amongst the weakly magnetic itinerant-electron magnets the cubic $B20$ compound MnSi has so far probably been studied most extensively. The interest originates thereby in a well defined hierarchy of energy scales comprising ferromagnetic exchange on the strongest scale, followed by Dzyaloshinsky-Moriya interactions on an intermediate scale and higher-order crystal-electric fields on the weakest scale. Focusing mainly on the highest energy scale, MnSi was studied extensively in the spirit of an essentially ferromagnetic material. These studies motivated in particular a detailed investigation of the suppression of the magnetic transition temperature under hydrostatic pressure, for which a marginal breakdown of the standard model of the metallic state, Fermi liquid theory, was anticipated.

In contrast, while hydrostatic pressure experiments revealed a suppression of magnetic order at $p_{\rm c} = 14.6$\,kbar, the transition was not continuous as expected but rather displayed a first-order character. Despite this lack of (second-order) quantum criticality,\cite{2007:Pfleiderer:Science} the low-temperature resistivity of MnSi abruptly changes from the $T^{2}$ Fermi liquid dependence for $p < p_{\rm c}$ to a stable $T^{3/2}$ non-Fermi liquid (NFL) behavior for $p > p_{\rm c}$.\cite{2001:Pfleiderer:Nature} This NFL regime persists over a remarkably wide range in temperature, pressure, and field and is accompanied by a partial magnetic order on timescales between $10^{-10}$\,s and $10^{-11}$\,s for $p_{\rm c} < p < p_{0} \approx 21$\,kbar.\cite{2004:Pfleiderer:Nature, 2007:Uemura:NaturePhys} Taken together, these findings clearly indicate the importance of the weak spin-orbit interactions in MnSi intimately connecting the NFL behavior to the magnetic structure which indeed is ferromagnetic only on short length scales, i.e., for large momenta.

The origin of this mysterious NFL resistivity appears to be related to the specific form of spin-orbit coupling and magnetic anisotropies mentioned above, which generate a long wavelength helimagnetic modulation $\lambda_{\textrm{h}} \approx 180$\,\AA\, along the crystalline $\langle111\rangle$ axes. An important consequence of the underlying hierarchy of energies concerns  the onset of magnetic order at ambient pressure and $T_{\rm C} \approx 29$\,K, which may be described within a Brazovskii scenario.\cite{1975:Brazovskii:SovPhysJETP} If $T_{\rm C}$ is approached from the mean-field-disordered paramagnet at high temperatures, isotropic chiral fluctuations evolve on a sphere in reciprocal space with a radius corresponding to the helical pitch. As the temperature is decreased further these fluctuations start to interact which gives rise to a fluctuation-disordered regime and finally triggers a fluctuation-induced first-order transition at $T_{\rm C}$.\cite{2013:Janoschek:PhysRevB} In a magnetic field the fluctuations are quenched and the phase transition changes to conventional second order at a field-induced tricritical point.\cite{2013:Bauer:PhysRevLett}

Finally, in recent years the largest scientific interest  MnSi and its isostructural siblings have attracted was related to the discovery of the Skyrmion lattice, i.e., a regular arrangement of topologically non-trivial spin whirls, in a small phase pocket just below $T_{\rm C}$.\cite{2009:Muhlbauer:Science, 2010:Munzer:PhysRevB, 2010:Pfleiderer:JPhysCondensMatter, 2010:Yu:Nature, 2011:Yu:NatureMater, 2012:Seki:Science, 2012:Adams:PhysRevLett, 2013:Milde:Science} If current is passed through this magnetic texture, the spins of the conduction electrons adiabatically follow the local magnetization thereby collecting a Berry phase. The latter may be interpreted as an emergent magnetic field of the flux quantum that each Skyrmion carries and leads to a topological contribution to the Hall effect.\cite{2009:Neubauer:PhysRevLett} In combination with the weak collective defect pinning of the Skyrmion lattice,\cite{2011:Adams:PhysRevLett} this very efficient coupling allows for spin transfer torque effects already at ultra-low current densities.\cite{2010:Jonietz:Science, 2012:Schulz:NaturePhys, 2012:Yu:NatCommun} A careful study of the evolution of the topological Hall effect as a function of pressure suggests an intimate connection of the Skyrmion lattice at ambient pressure with the NFL regime at high pressure.\cite{2013:Ritz:PhysRevB, 2013:Ritz:Nature} While this finding clearly suggests the presence of topologically non-trivial spin textures above $p_{\rm c}$, the microscopic origin of the stable $T^{3/2}$ dependence of the resistivity remains puzzling.

In order to unravel the underlying mechanisms that cause the NFL resistivity at high pressures, however, a full account for the conduction at ambient pressure is an important prerequisite. In fact, due to the complex Fermi surface of MnSi multiple bands are expected to contribute which requires a probe of the electrical conductivity that circumvents the limitations of conventional electric transport measurements and separately accounts for these different contributions.

A further important aspect concerns the coupling of the conduction electrons to the lattice since a change of the lattice constant by less than a percent, caused by either pressure\cite{2007:Pfleiderer:Science} or substitutional doping\cite{2004:Manyala:NatureMater, 2010:Bauer:PhysRevB}, is already sufficient to suppress magnetic order and distinctively change the electronic properties.
A substantial impact of the conduction electrons on the lattice was also found in ultrasonic attenuation~\cite{Kusaka:1976,Petrova:2009} and thermal expansion measurements, where an influence of magnetism can be detected up to 200\,K~\cite{Fawcett:1970,Matsunaga:1982,Stishov:2008}.

These issues may be addressed via Raman spectroscopy, i.e., inelastic light scattering, for which selection rules allow the separate detection of excitations having different symmetries. An improved optical setup for the first time permits us to collect enough inelastically scattered photons to study MnSi at low temperatures and in an external magnetic field. As will be illustrated in the following, the Raman response may be separated: (i) A lattice contribution which is particularly informative here since all optical phonons are Raman active as shown for the iso-structural compound FeSi~\cite{Nyhus:1995,Racu:2007}. In particular, also the infrared (IR) active polar phonons can be observed by light scattering due to the missing inversion symmetry in MnSi. The phonons are related to magneto-volume effects via the Gr\"uneisen parameter. Beyond that they can be used as microscopic probes providing information on possible spin-lattice interactions which may have eluded macroscopic thermodynamic measurements. (ii) A particle-hole continuum, where the pure symmetries correspond to different regions in the Brillouin zone~\cite{Devereaux:1994,Devereaux:2007}. The electronic continuum provides access to the anisotropy of fluctuations~\cite{Caprara:2005,Caprara:2011,Eiter:2013} and to the {\bf k}-dependence of the (two-particle) electronic relaxation rates.

To the best of our knowledge, there is only one Raman study carried out on MnSi prior to our work. Tite \textit{et al.} reported Raman data for elevated temperatures resolving eight out of nine predicted phonons~\cite{Tite:2010}. IR spectroscopy investigations were performed at low temperatures by Mena \textit{et al.}~\cite{Mena:2003}. Here, the authors studied the frequency and temperature dependence of the optical conductivity and the effective mass of the charge carriers revealing an optical conductivity below $T_{\rm C}$ that may not be described by the standard Drude formalism but with a phenomenological approach. The scattering rate and the Kramers-Kronig related effective mass thus deviate from Fermi liquid behavior which is surprising as it contradicts the results of conventional transport experiments. Yet, if the relaxation depends on momentum {\bf k}, as expected for MnSi~\cite{Belitz:2006}, IR spectroscopy yields averages over the entire Fermi surface.

The paper is organized as follows. In Sec.\,\ref{sec:ExpMeth} we start with a short account for material-specific parameters, the low-temperature Raman setup, and further experimental details, where we refer to the appendix for additional information. Results on the phononic and the electronic part of the Raman spectra are shown in Sec.\,\ref{sec:phonon_data} and \ref{sec:continuum}, respectively. The temperature dependence of phonon frequencies and linewidths as well as on the electronic carrier properties will be discussed in Sec.\,\ref{sec:Discussion} before we finally summarize our findings.

\section{Experimental methods}
\label{sec:ExpMeth}

\subsection{Sample Preparation}
\label{sec:samples}
MnSi crystallizes in the non-centrosymmetric cubic $B20$ ($P2_13$) structure. Large high-quality single crystals were prepared in a ultra-high vacuum compatible image furnace.\cite{Neubauer:2011,Bauer:2010} With optimized parameters the residual resistivity ratio (RRR) of the single-crystals grown can be as high as 300. Further details about the preparation and the characterization can be found in Ref.~\onlinecite{Bauer:2010}. The sample used for the Raman experiments had a RRR of approximately 100. An extensive study of the thermal conductivity as measured on samples from the same batch (cf.\,Fig.\,\ref{fig:thermal_conductivity}) will be published elsewhere. At 532.3\,nm and room temperature the complex index of refraction was determined to be $\hat{n} = 3.28 + 2.36\,i$. Using $\hat{n}$, the fraction of the absorbed photons and their polarization inside the sample can be calculated~\cite{Prestel:2012}. Before being mounted in the cryostat the sample was oriented via Laue diffraction and cleaved.

\subsection{Raman experiment}
\label{sec:Raman}

The Raman spectra were collected with the sample mounted in a bespoke He-flow cryostat permitting studies at temperatures between 1.8 and 330\,K. For the measurements the sample was inserted through a load-lock chamber into the He gas atmosphere of the variable temperature insert. The sample was thereby positioned in the center of a superconducting solenoid providing a magnetic field of up to 8\,T.

As a light source a solid state laser emitting at 532.3\,nm was used. A custom made objective lens collected the scattered light. The objective lens covered a solid angle of 0.37\,sr (numerical aperture N.A.~= 0.34) and corrected geometrical aberrations introduced by the cryostat windows. Extreme care was taken to control any extrinsic background. Thus the scattered light had to be spatially filtered. At the CCD detector approximately 1\,count\:$(\second \cdot \milli\watt)^{-1}$ arrived in the maximum of the strongest phonon peak and less than 0.1\,\cps in the electronic continuum as can be seen in the spectra shown in Fig.\,\ref{fig:raw-data}. For one spectrum six individual exposures of 300\,s each were added. This integration time was a tradeoff between the probability of cosmic spikes and the readout noise of the detector. For a total exposure of 1800\,s and an absorbed laser power $P_{\rm abs}=4$\,mW approximately 200 photo-electrons per point were collected in the electronic continuum of the spectra corresponding to a statistical error of 7\%. Further details are described in the appendix.

\subsection{Temperature determination}

The physical properties observed in the light scattering experiment corresponded to the temperature in the illuminated spot $T_{\rm{s}}$ which, due to local heating by the absorbed laser light, was higher than the temperature $T_{\rm h}$ of the sample holder by $\Delta T (T_{\rm h})$. In MnSi the standard methods to determine the laser heating fail due to the low scattering intensity and the lack of strongly temperature dependent features in the spectra throughout the whole temperature range. To circumvent this problem the laser heating was determined using a method that was developed for V$_3$Si\,\cite{Hackl:1983} and then applied to Nb$_3$Sn\,\cite{Hackl:1987}. The application to other (cubic) materials is possible since the transmission coefficient for light ${\cal T}=1-{\cal R}$ and the thermal conductivity $\lambda(T)$ are the only parameters relevant for the determination of $\Delta T (T_{\rm h})$. For more details see Appendix~\ref{sec:T-determination}. At the selected temperatures of the sample holder $T_{\rm{h}} =\,$2, 30, and 300$\,\kelvin$, $\Delta T$ is 2.9, 1.3, and 0.4$\,\kelvin$, respectively, for an absorbed laser power $P_{\rm abs}=4$\,mW. All measurements shown below are corrected for the laser heating. To simplify the notation, the temperature in the laser spot $T_{\rm{s}}$ is referred to as $T$ in the following.

\subsection{Response and selection rules}
\label{sec:pol}

Raman active excitations have spectral and symmetry properties. The measured spectra are proportional to the van Hove function $S_{\rm{i,s}}({{\bf q}\approx 0,\omega})$ with i and s indicating the polarizations of the incoming and scattered photons, respectively. Upon division by the thermal Bose factor $\{1+n(\omega,T)\} = (1 - e^{-\hbar\omega / k_{\rm{B}}T})^{-1}$ the Raman response $R\chi_{\rm{i,s}}^{\prime \prime}$ is obtained, where $R$ is a constant that absorbs experimental factors and takes care of the units.

The light polarizations $\hat{e}_{\rm{i,s}}$ are the key to access the symmetry properties of the excitations. The symmetry provides information on, e.g., phonon eigenvectors, collective modes or spin excitations and allow selective access to electron momenta in the Brillouin zone in the case of electron-hole excitations.\cite{Devereaux:1994,Devereaux:2007} $\hat{e}_{\rm{i}}$ and $\hat{e}_{\rm{s}}$ are set outside the cryostat such that the light polarization inside the sample is aligned with respect to the crystalline axes (cf. Appendix~\ref{Asec:exp}). $x$ and $y$ denote linear polarizations along the crystallographic $[100]$ and $[010]$ axes, respectively. $x'$ and $y'$ are rotated by $45\degree$ and point along $[110]$ and $[\bar{1}10]$, respectively. Left and right circularly polarized photons are represented as $l$ and $r$. The measurement configuration is given in Porto notation $(\hat{e}_{\rm{i}}\,\hat{e}_{\rm{s}})$.
In all cases a linear combination of symmetries is projected out. For the cubic space group of MnSi ($P2_1 3$) one obtains:
\begin{equation}
\label{eq:symmetries}
\begin{array}{c@{\:=\:}l c@{\:=\:}l}
  xx   & A_1 + \frac{4}{3}\, E                              & xy   & T_1 + T_2 \\[0.8ex]
  x'x' & A_1 + \frac{1}{3}\, E + T_2~~~~~~~~                & x'y' & E + T_1 \\[0.8ex]
  rr   & A_1 + \frac{1}{3}\, E + T_1                        & rl   & E + T_2
\end{array}
\end{equation}
The $A_1$, $T_1$, and $T_2$ symmetries are each fully projected in 3 polarizations. The $E$ symmetry appears in 5 polarizations. Factor group analysis predicts nine Raman active phonons in MnSi ($2 A_1 + 2 E + 5 T_2$)~\cite{Racu:2007}. The response of a symmetry $\mu$ can be determined via linear combinations of measured spectra
\begin{equation}
\label{eq:lincomb}
\begin{array}{l@{\:=\:}l}
  A_1 & \frac{1}{3}[(xx + x'x' + rr) - (x'y' + rl)]\\[0.8ex]
  E   & \frac{1}{3}[(xx + x'y' + rl) - \frac{1}{2}  (xy + x'x' + rr)] \\[0.8ex]
  T_1 & \frac{1}{3}[(xy + x'y' + rr) - \frac{1}{2}  (xx + x'x' + rl)] \\[0.8ex]
  T_2 & \frac{1}{3}[(xy + x'x' + rl) - \frac{1}{2}  (xx + x'y' + rr)].
\end{array}
\end{equation}
If the two spectra of each line in Eq.~\eqref{eq:symmetries} are added the entire response is obtained. The 3 sums should return exactly the same results and can be used to check the consistency of the measurements as shown, e.g., in Fig.\,\ref{fig:phonon-symmetry}\,(b). This consistency check was performed at 17, 36, 288 and 311\,K. At other temperatures only $xy$, $x'y'$, $rr$ and $rl$ spectra were measured. Four spectra are still sufficient to calculate the pure symmetries, e.g. $E=(x'y' + rl - xy)/2$, but are not enough for the full consistency check. In magnetic field only the circular polarizations were measured, because they are not affected by polarization rotations due to the Faraday effect.


\section{Results}
We studied electronic and lattice excitations of MnSi in the temperature range $1.8 < T_{\rm{h}} < 310$\,K as a function of the light polarizations. The main emphasis was placed on the temperature range below 50\,K close to the helimagnetic transition at $T_{\rm C} = 29$\,K. In some measurements a magnetic field of 4\,T was applied to suppress the helimagnetic modulation.

\begin{figure}
\includegraphics[width=0.44\textwidth]{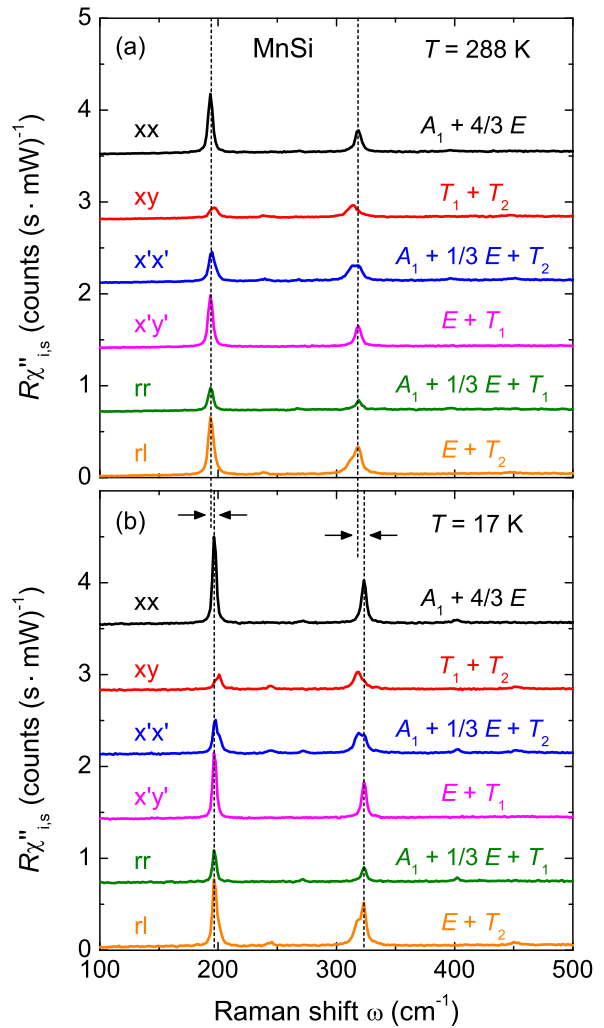}%
\caption{Raman spectra of MnSi at (a) 288 and (b) 17\,\kelvin. Plotted is the Raman susceptibility $R\chi^{\prime \prime}_{\rm{i,s}}(\omega,\,T)$ as a function of the energy shift $\omega$. The spectra are measured with different sets of light polarizations with respect to the crystallographic axes as explained in the text. For each polarization combination the symmetry components are indicated. For clarity the spectra are consecutively offset vertically by 0.7\,\cps. To point out the small frequency differences between $E$ and $T_2$ phonons, the positions of the two $E$ phonons in the $xx$ spectra are marked by dashed lines. Whenever both $E$ and $T_2$ excitations contribute to a spectrum, double peak structures appear. Upon lowering the temperature, all phonons harden. The black horizontal arrows indicate these frequency shifts for the $E$ phonons.
}
\label{fig:raw-data}
\end{figure}

In Fig.\,\ref{fig:raw-data} Raman spectra $R\chi_{\rm i,s}^{\prime \prime}$ at $288$ and $17\,\kelvin$ are shown. The results for the various polarization combinations are offset by 0.7\,\cps each. Along with the measurements the linear combination of symmetry components contributing to the spectra are indicated. All spectra consist of sharp peaks, originating from lattice excitations, and a very weak continuum arising from electron-hole or other excitations having a broad spectrum.

\begin{figure}
\includegraphics[width=0.44\textwidth]{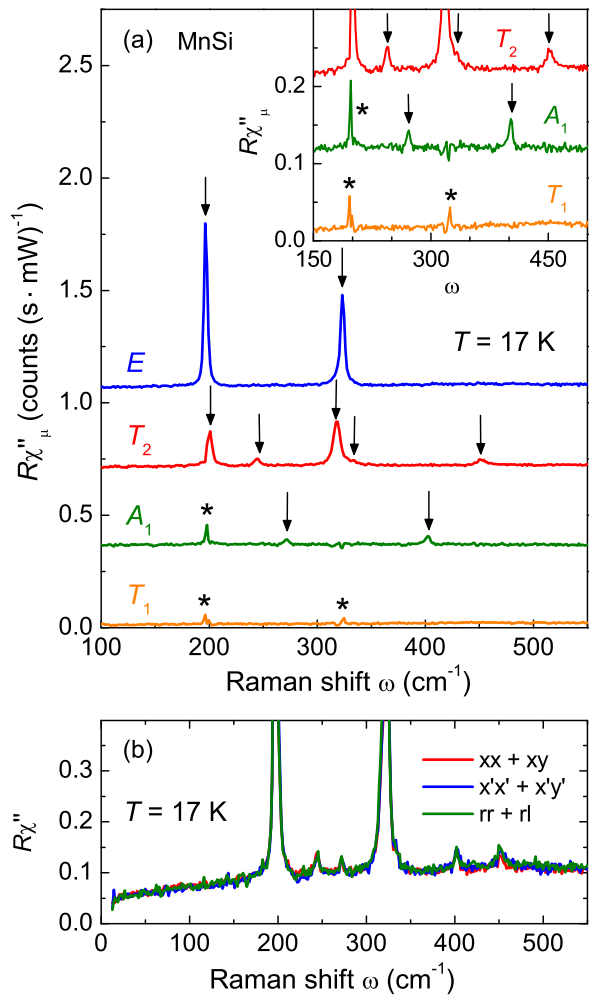}%
\caption{Symmetry resolved spectra of MnSi at 17\,\kelvin. (a) Pure symmetries can be obtained via linear combinations of Raman spectra measured with different polarization settings according to Eq.~(\ref{eq:lincomb}). For clarity the spectra are offset by 0.35\,\cps. All Raman active phonons ($2E + 5T_2 + 2A$) predicted by factor group analysis \cite{Racu:2007} are observed (marked by arrows). In the $A_1$ and $T_1$ spectra, there are additional peaks due to polarization leakage (marked by stars). The inset shows the weaker phonons of $T_2$ and $A_1$ symmetry on an expanded intensity scale (spectra offset by 0.1\,\cps). The $T_1$ spectrum contains the chiral excitations. It is featureless except for polarization leakage (stars), but the intensity of the continuum is in the same order of magnitude as the other symmetries. (b) Sums of spectra having orthogonal polarizations of the scattered photons. Two orthogonal measurements cover the full response of the sample, thus their sums must be invariant. This is used to check the consistency of the measurements.
}
\label{fig:phonon-symmetry}
\end{figure}

The pure symmetry components can be derived from the experimental spectra (Fig.~\ref{fig:raw-data}) via the linear combinations compiled in Eq.~(\ref{eq:lincomb}) and are shown in Fig.~\ref{fig:phonon-symmetry}\,(a) for low temperatures. Fig.~\ref{fig:phonon-symmetry}\,(b) demonstrates that the six spectra shown in Fig.~\ref{fig:raw-data}\,(b) are consistent. This holds for both the phonons and the continuum.

\subsection{Phonons}
\label{sec:phonon_data}
In the raw data in Fig.\,\ref{fig:raw-data} not all phonons are immediately observable since some of them are rather close in energy. However, the temperature dependence of the four strongest lines can be read directly from the data: Two in $E$ (marked by dashed lines) and two in $T_2$ symmetry. The energies of the two strong $T_2$ lines are only slightly different from those of the $E$ phonons. Therefore, the $E$ and $T_2$ modes in the $x'x'$ and $rl$ spectra appear as double peak structures. However, they can be observed independently in the $x'y'$ and $xy$ spectra, respectively. There are also three weak $T_2$ and two weak $A_1$ phonon lines. These low intensity phonons are hardly visible on the scale of Fig.\,\ref{fig:raw-data}, but are nonetheless marked in Fig.\,\ref{fig:phonon-symmetry}. As expected, no phonons are present in $T_1$ symmetry.

\begin{table}
\begin{ruledtabular}
\begin{tabular*}{0.45\textwidth}{lccc}
        &    \multicolumn{3}{ c }{Phonon frequency $\omega_{\rm{ph}}$~$[\rm{cm}^{-1}]$} \\
        & $T \approx 295\,\kelvin$      & $T = 288\,\kelvin$  & $T = 17\,\kelvin$       \\
        & (Ref.~\onlinecite{Tite:2010}) & (this\;work)        & (this\;work)            \\[0.2ex]
  \hline\\[-2ex]
  $A_1$ & 268                           & 268                 & 271                     \\
  $A_1$ & 398                           & 396                 & 402                     \\
  $E$   & 193                           & \bf{193.5}          & \bf{196.9}              \\
  $E$   & 319                           & \bf{318.5}          & \bf{323.4}              \\
  $T_2$ & 194                           & \bf{196.3}          & \bf{200.9}              \\
  $T_2$ & 236                           & 239                 & 244                     \\
  $T_2$ & 316                           & \bf{313.9}          & \bf{318.6}              \\
  $T_2$ &  --                           &  --                 & 332                     \\
  $T_2$ & 448                           & 447                 & 452                     \\
\end{tabular*}
\caption{Energies of the Raman active phonon modes in MnSi. At low temperatures all nine predicted peaks are resolved. Room temperature measurements are shown together with the phonon positions derived in Ref.\,\onlinecite{Tite:2010}. The frequency values of the four intense phonons can be determined with an accuracy of about $\pm 0.2\,\centi\metre^{-1}$ using Voigt fits  (bold face). The other phonons are determined with a precision of $\omega_{\rm{ph}} \pm 1\,\centi\metre^{-1}$. \label{tab:phononFreq}}
\end{ruledtabular}
\end{table}

In Table~\ref{tab:phononFreq} we show the frequencies of the phonons measured at 17 and 288\,K along with the results of Tite \textit{et al.}~\cite{Tite:2010}. At room temperature, only eight of the nine expected phonons can be detected while all are seen at low temperature. The missing $T_2$ line turns out to be very weak and appears only as a shoulder at 332\,\wn on the high-energy side of the line at 318.6\,\wn. A comparison with the phonon positions in the isostructural compound FeSi\,\cite{Racu:2007} supports this finding.

Tite and coworkers\cite{Tite:2010} performed their measurements at elevated temperatures using various high laser powers. Considering their laser power $P_{\rm abs}$ and the focus size $r_0$ in Eq.~\eqref{eq:heating} (Appendix), the associated temperatures may be estimated to be as high as 650\,K. The huge heating $\Delta T$ implies an inhomogeneous broadening of the phonon lines explaining the different line widths reported in Ref.~\onlinecite{Tite:2010} and in our study. On the other hand, if the temperature dependence of the phonon energies $\omega_{\rm{ph}}(T)$ is assumed to be linear above room temperature, Tite \textit{et al.} are able to extrapolate the phonon positions down to 295\,K in satisfactory agreement with our data (cf. Table~\ref{tab:phononFreq}).

In our study the temperature range was extended down to 4\,K. Particular emphasis was placed on the temperature range around the helimagetic phase transition at $29\,\kelvin$. With decreasing temperature all phonons shift to higher energies. At the transition this trend is reversed, and the phonons soften by typically half a percent. To verify these small shifts in the phonon frequencies, the stability of the experimental setup was rechecked before and after each measurement with the spectral lines of a Neon calibration lamp. Fitting Voigt profiles to the phonon lines allowed us to reproducibly determine the frequencies and widths with an accuracy of approximately $\pm 0.2\,\centi\metre^{-1}$. A Voigt fit consists of a convolution of a Lorentzian and a Gaussian profile. The Gaussian width was fixed at 2\,\wn and represents the spectral resolution. The width, position, and spectral weight of the Lorentzians were the fitting parameters. In this way better results could be obtained than with pure Lorentzian lineshapes since the phonons are rather narrow, and the spectral resolution cannot be reduced any further for intensity reasons. The position of the phonon lines was determined from raw data rather than from pure symmetry spectra to avoid artifacts arising from linear combinations of the spectra (cf. Fig.\,\ref{fig:phonon-symmetry}). Also the $E$ and $T_2$ phonons that are very close in frequency can be separated by using $x'y'$ and $xy$ measurements~[cf. Eq.~\eqref{eq:symmetries}]. The results for the temperature dependence of the position and line width of these phonons are shown in Figs.\,\ref{fig:phonon-E} and~\ref{fig:phonon-T2} in section~\ref{sec:phonons} and will be discussed there in detail.

The $xy$ spectrum projects both $T$ symmetries. In $T_1$ symmetry neither phonons nor conduction electrons appear, but only excitations where the chiral symmetry changes. These excitations include transitions between states with well defined orbital character such as crystal field excitations\cite{Hayes:1978}.
As shown in Fig.~\ref{fig:phonon-symmetry}\,(a), displaying the pure symmetries, the response does not vanish indicating the presence of excitations beyond scattering from phonons and electron-hole pairs that will be addressed in section \ref{sec:continuum} and \ref{sec:electronicRaman}. In addition, the inset of Fig.~\ref{fig:phonon-symmetry}\,(a) demonstrates that the $T_1$ continuum is too strong to be neglected. Therefore, if the continua of the other symmetries are to be analyzed, only the pure symmetries [Fig.~\ref{fig:phonon-symmetry}\,(a)] can be used. Prior to an analysis of the continuum it is necessary to subtract the phonons. To this end we apply the fitting procedure described above to all phonons, subtract them and thereby obtain the isolated continuum.

\subsection{Continuum}
\label{sec:continuum}

\begin{figure}
\includegraphics[width=0.44\textwidth]{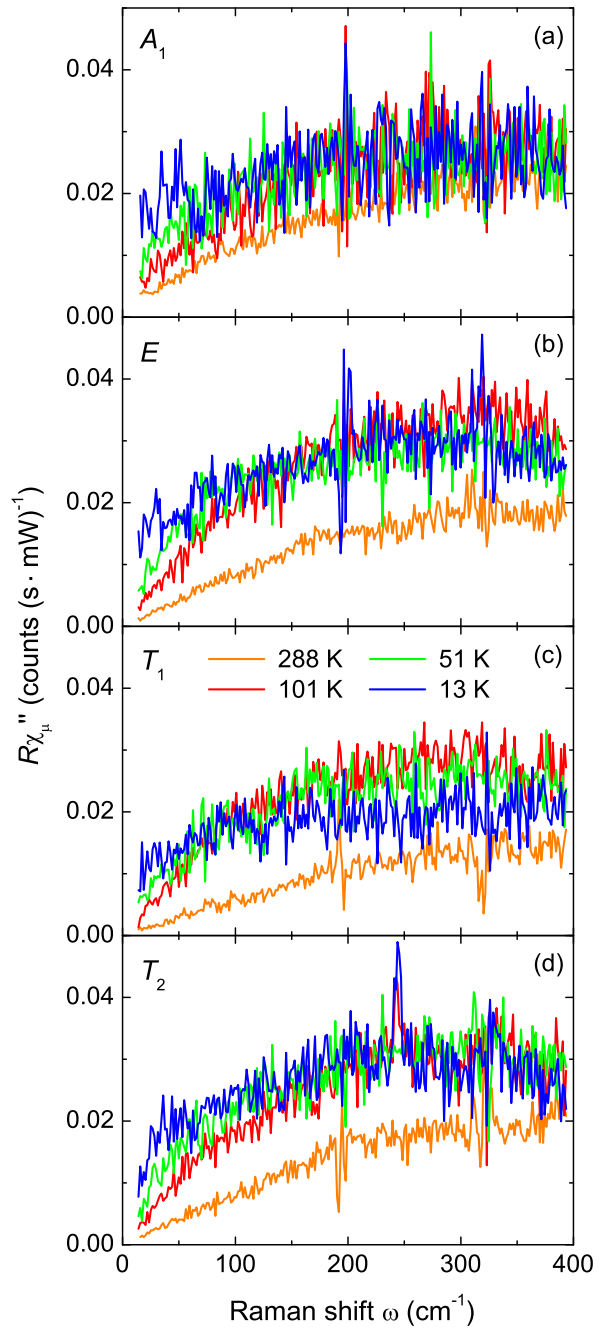}%
\caption{Temperature and symmetry dependence of the electronic continua in all pure symmetries~$\mu = A_1,\, E,\, T_1,\, T_2$. The response is obtained via linear combinations of $xy$, $x'y'$, $rr$ and $rl$ spectra. Phonons were fitted with Voigt profiles and subtracted.
}
\label{fig:continua}
\end{figure}

The electronic continua observed in the pure symmetries are plotted in Fig.~\ref{fig:continua} for various temperatures in the range $13\le T\le 288$\,K. The continua were obtained by subtracting the individually fitted phonons from the spectra. Although the fits were reproducible and stable there are occasionally remainders of phonons after the subtraction, particularly in the low temperature spectra having phonon line widths similar to the spectral resolution. The data are relatively noisy in general since the scattering cross sections are low (see section~\ref{sec:Raman}) and the spectra in pure symmetries are the result of subtraction procedures [cf. Eq.\,\eqref{eq:lincomb}].

In all symmetries, the spectra exhibit a substantial temperature dependence at low energies. The $E$, $T_1$ and $T_2$ spectra are also temperature dependent at higher energies. At 13\,K, the $A_1$, $E$, and $T_1$ spectra are too steep below 50\,\wn to allow the extrapolation to zero at $\omega \rightarrow 0$ to be observed. In fact, $\chi^{\prime\prime}(-\omega) = -\chi^{\prime\prime}(\omega)$ is expected since, for causality reasons, the response is antisymmetric. Only the $T_2$ spectra show the expected linear energy dependence at all temperatures and extrapolate approximately to zero. The variation with temperature implies a substantial increase of the initial slope of the response.

As discussed in detail by Opel and coworkers \cite{Opel:2000}, the inverse initial slope, $(\partial\chi^{\prime\prime}(\omega)/\partial\omega)^{-1}$ corresponds to a Raman resistivity and is therefore a useful quantity to be compared with transport measurements. It can be extracted in a similar fashion as conductivity from the optical spectra (IR) if a wider energy range is known. In the case of the Raman spectra the requirements as to the known spectral range are much more relaxed than in optical spectroscopy since the Kramers-Kronig transform converges rapidly~\cite{Opel:2000}. The results of this analysis, in particular those in the zero-energy limit, will be shown and discussed in section~\ref{sec:electronicRaman} (Figs.~\ref{fig:Gamma}, \ref{fig:lambda} and \ref{fig:transport}).

\section{Discussion}
\label{sec:Discussion}

Major anomalies of the lattice and the carrier properties are observed close to \Tc while the high temperature range develops more conventionally. First we discuss the phonons, then the electronic response. In each case, we start with the behavior above the helimagnetic transition at \Tc followed by a detailed study close to \Tc.

\subsection{Temperature dependence of the phonons}
\label{sec:phonons}

\begin{figure}
\includegraphics[width=0.44\textwidth]{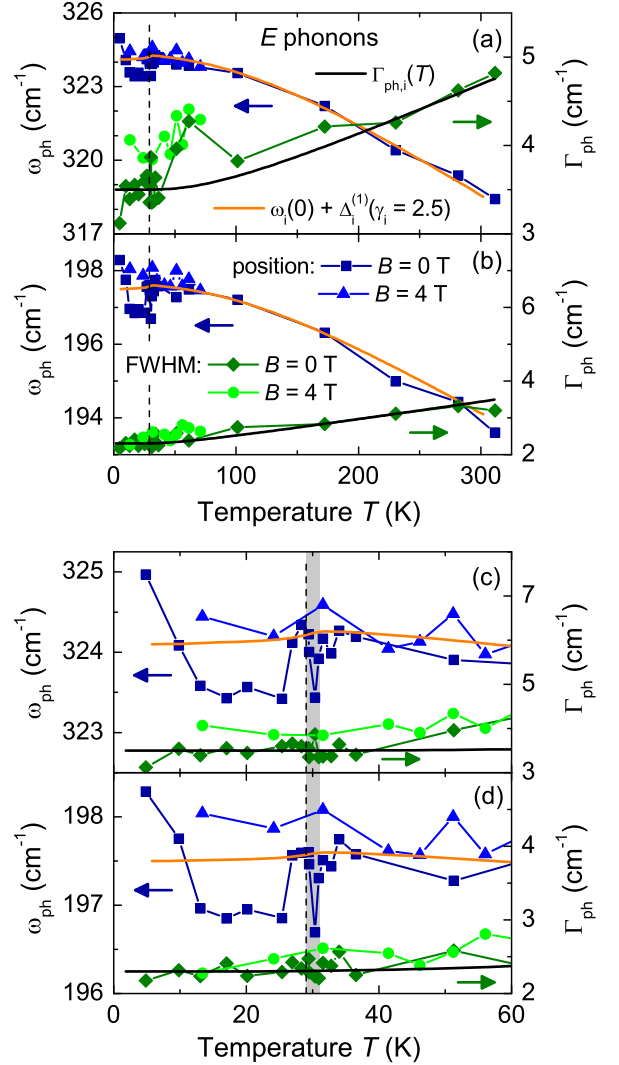}%
\caption{Temperature dependence of the $E$ phonons. Shown are the frequencies (left scale) and linewidths (right scale) of the two strongest lines (labeled by $E^{(318)}$ and $E^{(194)}$ according to their room temperature positions). (a) and (b) show the analysis for the full temperature range, (c) and (d) zoom in on low temperatures. For both phonons, $E^{(318)}$ and $E^{(194)}$, the linewidth can be described by the model of anharmonic decay~\cite{Klemens:1966} (black lines). The main contribution to the frequency change is due to thermal expansion and can be described by a constant Gr\"uneisen parameter $\gamma_i =2.5$ above 35\,K (orange line). Right above \Tc (dashed vertical line) in the fluctuation disordered regime~\cite{2013:Bauer:PhysRevLett} (shaded) there is a dip in the phonon frequency. If the helical order is suppressed by a magnetic field of $4\,\tesla$ (triangles and circles), the anomalies in the phonon frequencies disappear.
}
\label{fig:phonon-E}
\end{figure}

\begin{figure}
\includegraphics[width=0.44\textwidth]{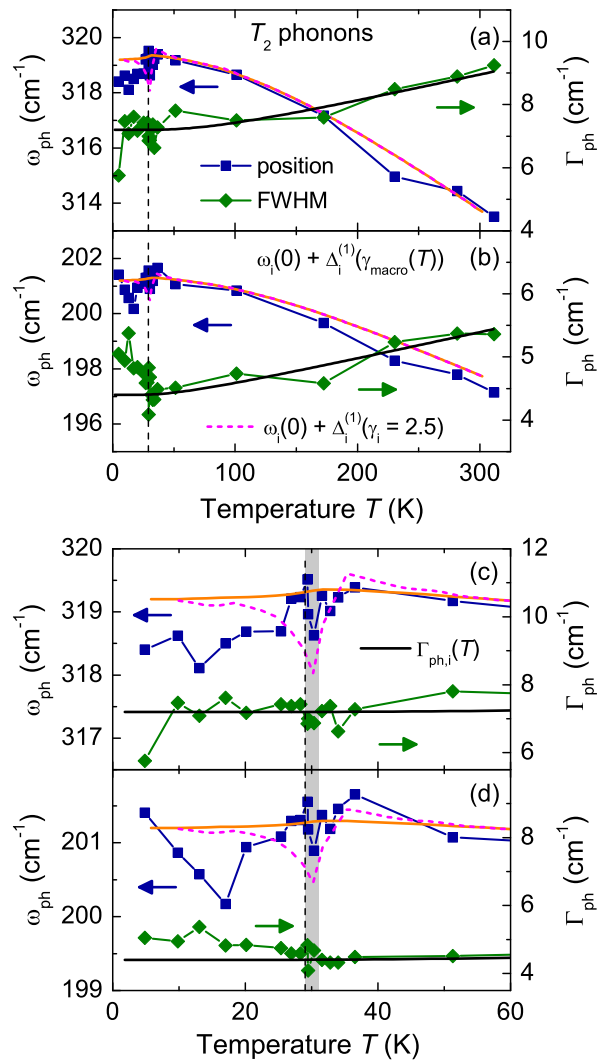}%
\caption{Temperature dependence of $T_2$ phonons. Shown are the frequencies (left scale) and linewidths (right scale) of the two strongest lines (labeled by $T_2^{(314)}$ and $T_2^{(197)}$ according to their room temperature positions). (a) and (b) show data for the full temperature range, (c) and (d) zoom in on low temperatures. The phonon width deviates from the predictions of the Klemens model~\cite{Klemens:1966} (black lines) only for the $T_2^{(197)}$ phonon below \Tc (dashed vertical line). Above 35\,K the frequency change of both phonons can be explained in terms of a thermal expansion shift $\Delta^{(1)}_i(T)$ (orange lines) assuming a constant $\gamma_i = 2.5$ for all phonon modes $i$. There is a dip in the phonon frequency right above \Tc in the fluctuation disordered region~\cite{Janoschek:2013,2013:Bauer:PhysRevLett} (shaded). The dip can be reproduced qualitatively if the macroscopic Gr\"uneisen parameter $\gamma_{\rm macro}(T)$ is inserted into Eq.~\eqref{eq:Delta1} as described in the text (dashed magenta line). In the helimagnetic phase the phonon frequencies are lower than those predicted by the thermal expansion.
}
\label{fig:phonon-T2}
\end{figure}

Figs.\,\ref{fig:phonon-E} and~\ref{fig:phonon-T2} show the temperature dependences of the frequencies and linewidths of the two strongest $T_2$ and $E$ phonons as derived from Voigt fits (cf. section~\ref{sec:phonon_data}). As temperature decreases in the range $T > 35\,\kelvin$, there is the typical blue shift and line narrowing. Right above \Tc, in the range between 35 and 29\,K, there is a dip in all phonon frequencies. Below the transition the phonons anomalously soften by approximately 0.5\,\wn as can be seen more clearly in panels (c) and (d) of Figs.\,\ref{fig:phonon-E} and~\ref{fig:phonon-T2}. Except for the high-energy $T_2$ mode all lines harden again below approximately $1/2\,T_{\rm C}$. The $T_2$ mode at 201\,\wn [Fig.\,\ref{fig:phonon-T2}\,(d)] reaches the same energy as found at \Tc whereas the $E$ lines (Fig.\,\ref{fig:phonon-E}) even exceed the low-temperature extrapolation value expected from the range $T>T_{\rm C}$.

All anomalies of the phonon energies vanish completely in a magnetic field of 4\,T (see Fig.\,\ref{fig:phonon-E}), which is well above $0.6\,\tesla$, the upper critical field of the helimagnetic modulation. Nevertheless, a crossover temperature $T_{\rm{cr}}$ continues to exists that separates the regimes dominated by either magnetism or temperature. For MnSi at $B = 4\,\tesla$, $T_{\rm{cr}}$ is approximately 40\,K.\cite{Demishev:2012} However, no anomalies in the phonon positions and widths are detected at this temperature suggesting that the phonon anomalies originate from the chiral order.

We first focus on the high-temperature part. In the harmonic approximation of lattice dynamics, the phonon frequencies are not temperature dependent, and the phonon lifetime is infinite. In real systems and to describe the experimentally observed temperature dependences of frequency and lifetime, anharmonic contributions to the lattice potential have to be taken into account. A consequence of these anharmonic terms are collisions between phonons which lead to changes in the phonon frequency and to the creation or annihilation of phonons resulting in a finite lifetime. Frequency shift and broadening can be described in terms of the real and imaginary part of the self energy\cite{Menendez:1984}
\begin{equation}
\label{eq:self_energy}
\Sigma_i(T) = \Delta_i(T) + \imath \Gamma_i(T)
\end{equation}
corresponding to the position and width of phonon $i$, respectively. The anharmonic effects can be treated perturbationally and were extensively studied by several authors~\cite{Klemens:1966,Postmus:1968,Borer:1971,Menendez:1984}.

Here we consider only optical phonons in the center of the Brillouin zone. Taking into account energy and momentum conservation, an optical phonon with $\vq=0$ decays into two acoustic phonons of opposite wave vectors $\omega_1(\vq,j_1)+\omega_2(-\vq,j_2) = \omega_{\rm{ph}}$. The indices $j_1$ and $j_2$ label acoustic phonon dispersion branches. Klemens assumes~\cite{Klemens:1966} that the most relevant decay channels are symmetric, $\omega_1(\vq,j_1) = \omega_2(-\vq,j_1) = \omega_{\rm{ph}}/2$, and within the same acoustic phonon branch, e.g. $j_1$. Then the temperature dependence of the linewidth of phonons reads
\begin{equation}
\label{eq:Klemens}
\Gamma_{{\rm ph},\,i}(T) = \Gamma_i(0) \Biggl[ 1 + \frac{2 \cdot \lambda_{p-p,i}}{\exp (\frac{\hbar \omega_i(0)}{2 k_{\rm{B}}T}) - 1} \Biggr].
\end{equation}
The width $\Gamma_i(0)$ and position $\omega_i(0)$ of the $i$-th Raman line for $T \rightarrow 0$ can be obtained by extrapolating $\Gamma_{{\rm ph},\,i}(T)$ to zero temperature from the range above \Tc before the anomalies set in. As temperature rises, the linewidth increases by two times the Bose factor at $\omega_i(0)/2$, representing the symmetric decay channel. $\lambda_{p-p,i}$ was introduced as the only fitting parameter and is interpreted as phonon-phonon coupling strength. %
More general calculations including asymmetric decay channels result in a better agreement with experiment in some semiconductors~\cite{Cowley:1965,Menendez:1984}. In MnSi, however, asymmetric phonon decays as well as four phonon processes turn out to be negligible, and the Klemens model provides a reasonable fit to the linewidth of all phonons studied, except $T_2^{(197)}$ at low temperatures.

The frequency shift of the peaks is described by the real part of the self energy $\Delta_i(T)$ in Eq.\,\eqref{eq:self_energy}. The temperature dependence of phonon $i$ reads
\begin{equation}
\label{eq:frequency_shift}
\omega_{{\rm ph},i}(T) = \omega_i(0) + \Delta_i(T).
\end{equation}
Here, only the two lowest-order contributions, $\Delta_i(T) = \Delta^{(1)}_i(T) + \Delta^{(2)}_i(T)$, will be discussed. A  detailed description may be found in Refs.\,\onlinecite{Menendez:1984} and \onlinecite{Cowley:1965}. The leading term $\Delta^{(1)}_i(T)$ originates from the thermal lattice expansion, hence is related to a macroscopic quantity. Before we derive $\Delta^{(1)}_i(T)$ we show that $\Delta^{(2)}_i(T)$ is small for the phonon energies and the temperature range studied here.

The second-order contribution $\Delta^{(2)}_i(T)$ results from the anharmonic decay of phonons. The approximate relationship between the eigen-frequency $\omega_i(0)$, the resonance frequency $\omega_{{\rm ph},i}$ (peak maximum) and the linewidth $\Gamma_i$ (FWHM), $\omega_{{\rm ph},i} = \sqrt{\omega_i(0)^2-\Gamma_i^2}$, of a damped harmonic oscillator along with Eq.\,\eqref{eq:Klemens} yield a shift
\begin{equation}
\label{eq:Delta2}
\Delta^{(2)}_i(T) = - C_i \Biggl[ 1 + \frac{4 \cdot \lambda_{p-p,i}}{\exp \frac{\hbar \omega_0}{2 k_{\rm{B}}T} - 1} \Biggr],
\end{equation}
with $C_i = \Gamma_i(0)^2/2\omega_i(0)$ and all other parameters as defined above. The parameters for the phonons analyzed here are compiled in Table\,\ref{tab:fitParameter}.
\begin{table}
\begin{ruledtabular}
\begin{tabular*}{0.45\textwidth}{lcccc}
                          \multicolumn{5}{ c }{Fit parameter}                                                \\
                                                    & $E^{(194)}$    & $E^{(318)}$    & $T_2^{(197)}$   & $T_2^{(314)}$\\[0.3ex]
  \hline\\[-2ex]
  $\omega_i(0)$~$[\rm{cm}^{-1}]$                    & 197.5          & 324.1          & 201.2           & 319.2       \\
  $\Gamma_i(0)$~$[\rm{cm}^{-1}]$                    & 2.3            & 3.5            & 4.4             & 7.2         \\
  $C_i=\Gamma_i(0)^2/2\omega_i(0)$~$[\rm{cm}^{-1}]$ & 0.013          & 0.019          & 0.048           & 0.081       \\
  $\lambda_{p-p,i}$                                 & 0.15           & 0.2            & 0.07            & 0.14        \\
\end{tabular*}
\caption{Phonon parameters. According to their symmetry and room temperature frequency, the phonons $i$ are labeled $E^{(194)}$, $E^{(318)}$, $T_2^{(197)}$, and $T_2^{(314)}$. $\omega_i(0)$, $\Gamma_i(0)$ are experimentally determined constants. The phonon-phonon coupling strength $\lambda_{p-p,i}$ results from the fit of the phonon width according to Eq.\,\eqref{eq:Klemens} and was used again in Eq.\,\eqref{eq:Delta2}.
\label{tab:fitParameter}}
\end{ruledtabular}
\end{table}
The additional factor 2 in the numerator compared to Eq.\,\eqref{eq:Klemens} is due to the damped harmonic oscillator approximation. For the phonon-phonon coupling $\lambda_{p-p,i}$, the values obtained from Eq.\,\eqref{eq:Klemens} are used again. It turns out that the coupling is stronger for $E$ than for $T_2$ phonons and also stronger for the high frequency modes.

As $\Gamma_i(0)$ is small in comparison to $\omega_i(0)$, the coefficient $C_i$ is small and the contribution of $\Delta^{(2)}_i$ to the phonon shift is at least two orders of magnitude smaller than the thermal expansion shift $\Delta^{(1)}_i$ and therefore negligible. $\Delta^{(2)}_i$ gives significant contributions only for phonons with very low frequencies or temperatures in excess of the Debye temperature $\Theta_{{\rm Debye}}$ being as high as $600\,\kelvin$ here~\cite{Zinoveva:1974,Jasperse:1966,Borer:1971}. Obviously, the dominating contributions to the widths and the frequency shifts of the phonons result from different mechanisms and thus are not directly interrelated. This explains why the anomalies in the phonon frequencies do not have a direct correspondence in the linewidths.

The first order term $\Delta^{(1)}_i$ depends on the unit cell volume. In general, the resonance frequencies of the phonons increase upon decreasing the unit cell volume since the forces between the atoms increase with decreasing distance. The change can be quantified via the microscopic Gr\"uneisen parameter~\cite{AshcroftMermin:1988} $\gamma_i$ of mode $i$ being defined as the negative logarithmic derivative of a normal-mode frequency $\omega_i$ with respect to the volume $V$,
\begin{equation}
\label{eq:grueneisenDef}
\gamma_i = - \partial (\ln \omega_i) / \partial (\ln V).
\end{equation}
The related thermodynamic quantity is the macroscopic Gr\"uneisen parameter $\gamma_{{\rm macro}}(T)$ which is a weighted average of the contributions from the lattice, as well as from charges and magnetism~\cite{Barron:1980}. For the phonon part it can be shown that the relative weight is given by their individual contributions to the specific heat~\cite{AshcroftMermin:1988}. While $\gamma_{{\rm macro}}(T)$  is approximately constant in conventional insulators, it may vary considerably with temperature in complex metallic systems whenever different contributions determine the thermodynamic properties~\cite{Barron:1980,White:1986}. $\gamma_{{\rm macro}}(T)$ can be determined from experimentally accessible thermodynamic properties alone~\cite{White:1986},
\begin{equation}
\label{eq:grueneisen(T)}
\gamma_{{\rm macro}}(T)=\frac{3 \cdot \alpha(T) \cdot K(T) \cdot V^{\rm mol}(T)}{C_p^{\rm mol}(T)}
\end{equation}
and can directly be calculated up to 100\,K using the published data of the coefficient of thermal expansion $\alpha(T)$, the bulk modulus $K(T)$, the molar volume $V^{\rm mol}(T)$, and the molar heat capacity  $C_p^{\rm mol}(T)$  \cite{Stishov:2008,Petrova:2009,Pauling:1948,2013:Bauer:PhysRevLett}. Each of the quantities contributing to $\gamma_{{\rm macro}}$ is temperature dependent with strong anomalies close to \Tc. Consequently, also $\gamma_{{\rm macro}}$ depends on temperature as shown in Fig.\,\ref{fig:Grueneisen} in the Appendix. For $T \rightarrow 0$, $\gamma_{{\rm macro}}(T)$ is expected to asymptotically approach zero as $\alpha$ vanishes. Upon approaching \Tc from low temperatures, Eq.\,\eqref{eq:grueneisen(T)} yields large negative values of approximately -15. Slightly above the transition there is a pronounced dip-hump structure with the minimum at 30\,K, the maximum at 35\,K and a sign change in between. Upon further increasing the temperature, $\gamma_{{\rm macro}}(T)$ asymptotically approaches the constant value of 2.5 from above. Therefore, $\gamma_{{\rm macro}}$ was set to 2.5 above 100\,K, because not all quantities entering Eq.~\eqref{eq:grueneisen(T)} were available up to room temperature.

In MnSi magnetostrictive effects contribute to the anomalies of $\gamma_{{\rm macro}}(T)$ around \Tc, and relatively large values even at elevated temperatures are to be expected. For instance, the magnetic contributions to $\alpha(T)$ play an important role up to at least 200\,K and are of the same order of magnitude as the non-magnetic ones.\cite{Matsunaga:1982} At low temperatures magnetic order drives $\alpha(T)$ even negative, i.e. the lattice expands upon cooling, and leads to the strong dip of $\gamma_{{\rm macro}}(T)$ around \Tc~\cite{Matsunaga:1982,Stishov:2008,Petrova:2009}.

It is not the purpose of this study to systematically disentangle the various contributions to $\gamma_{{\rm macro}}(T)$ or to determine their respective weight. Rather, we wish to find out to which extent $\Delta^{(1)}_i(T)$ can be understood in terms of bulk properties and where microscopic effects can be pinned down. We first calculate $\Delta^{(1)}_i(T)$ using Eq.\,\eqref{eq:grueneisenDef}.
For constant $\gamma_i$, Eq.\,\eqref{eq:grueneisenDef} can be integrated~\cite{Postmus:1968} yielding an expression for the frequency shift $\Delta^{(1)}_i(T)$ of phonon $i$,
\begin{equation}
\label{eq:Delta1}
\Delta^{(1)}_i(T) = \omega_i(0) \left\{ \exp{\left[-3\gamma_{i} \int_{0}^{T} \alpha(T^{\prime}) dT^{\prime}\right]}-1 \right\}.
\end{equation}
The phonon frequency $\omega_i(0)$ is the only free parameter which can be determined for each branch $i$ by a fit to the high-temperature data. The temperature dependences of $\omega_{{\rm ph},i}(T)$ according to Eq.\,\eqref{eq:Delta1} are plotted as orange lines in Figs.\,\ref{fig:phonon-E} and~\ref{fig:phonon-T2}. At temperatures above 100\,K the frequency changes of the Raman phonons are well described by $\Delta^{(1)}_i(T)$ with a constant Gr\"uneisen parameter $\gamma_i = 2.5$ in good agreement with the asymptotic limit of $\gamma_{{\rm macro}}(T)$. Approaching the phase transition from above, the phonon frequencies increase as does $\gamma_{{\rm macro}}(T)$ (see also Fig.~\ref{fig:Grueneisen} in the Appendix). Similarly, the experimentally observed dip at 30\,K\,$>$\,\Tc has a corresponding anomaly in $\gamma_{{\rm macro}}(T)$. At \Tc all phonon frequencies jump back to the value at $T>30$\,K and then soften again. The phonon anomaly at \Tc is unparalleled in any of the macroscopic quantities. While the $T_2^{(314)}$ phonon seems to stay at lower frequency, the other phonons reach frequencies above $\omega_i(0)$. The prediction according to Eq.~\eqref{eq:Delta1} shows only a tiny kink which originates from $\alpha(T)$ (since all other quantities in Eq.\,\eqref{eq:Delta1} are constant). For $T \rightarrow 0$, $\Delta^{(1)}_i(T)$ vanishes as $\int \! \alpha \, dT^{\prime}$ in the exponent goes to zero.

\begin{figure}
\includegraphics[width=0.44\textwidth]{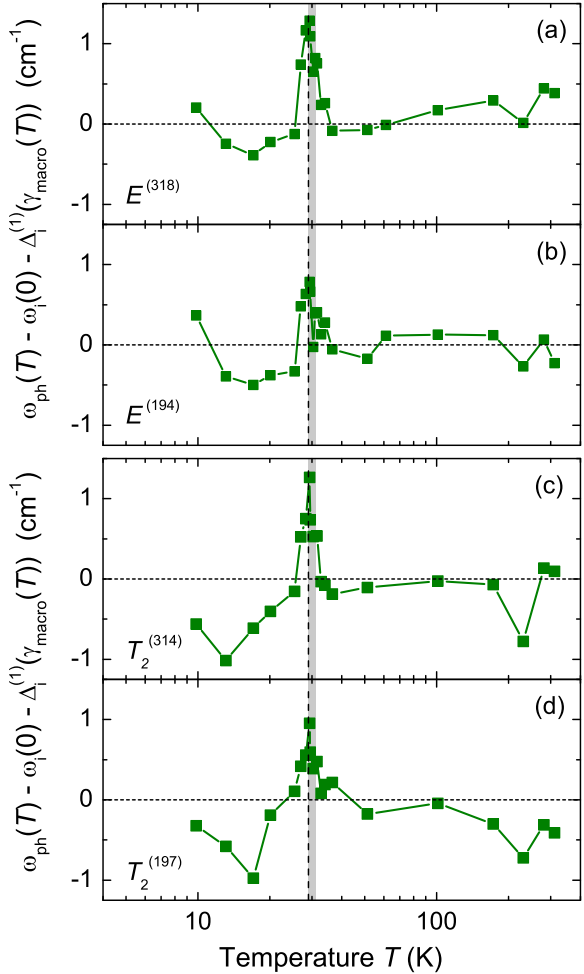}%
\caption{Difference of the experimental phonon energies $\omega_{\rm ph}(T)$ and those calculated via Eq.\,\eqref{eq:frequency_shift} and~\eqref{eq:Delta1} using $\gamma_{{\rm macro}}(T)$. Note the logarithmic temperature scale. The anomalies above \Tc vanish almost completely while those at \Tc have no correspondence in the thermodynamical properties.
}
\label{fig:differences}
\end{figure}

To summarize this part, we find that the widths of the four strongest phonons can be well understood in terms of symmetric anharmonic decay in the entire temperature range studied. The phonon energies are compatible with the thermodynamic properties in the temperature range above 100\,K. Below 100\,K the temperature dependence of the phonon lines demonstrate that microscopic and macroscopic properties develop individually. While the anomaly right above \Tc is still clearly visibly in either case the magnetism seems to have a quite specific influence on the phonons which is not directly visible in the thermodynamic properties. The results with applied field demonstrate that there are no phonon anomalies without chiral order.

In order to disentangle thermodynamic and microscopic properties we insert $\gamma_{{\rm macro}}$ in Eq.\,\eqref{eq:Delta1} and recalculate $\Delta^{(1)}_i(T)$. This is motivated by the proximity of the anomalies in the phonon energies $\omega_{{\rm ph},i}(T)$ and of $\gamma_{{\rm macro}}(T)$ but cannot be justified mathematically since Eq.\,\eqref{eq:grueneisenDef} yields Eq.\,\eqref{eq:Delta1} only for a constant $\gamma_i$. As shown in Fig.\,\ref{fig:phonon-T2} the shift obtained in this way (dashed magenta lines) is identical to that for $\gamma_i$ (orange lines) down to approximately 50\,K but deviates below. The anomaly of $\omega_{{\rm ph},i}(T)$ observed right above \Tc has now a correspondence in the prediction while that at \Tc cannot be reproduced. Similar results are found for $E$ symmetry but are not plotted to avoid overloading Fig.\,\ref{fig:phonon-E}.

The interrelation of microscopic and thermodynamic properties can be visualized in a better way by looking at the difference between the experimental frequencies $\omega_{{\rm ph},i}(T)$ and those calculated on the basis of Eq.\,\eqref{eq:Delta1} using $\gamma_{{\rm macro}}(T)$. Fig.\,\ref{fig:differences} demonstrates that the anomaly above \Tc vanishes almost completely (with small phonon-specific variations) while that at \Tc is rather pronounced. Although the use of $\gamma_{{\rm macro}}(T)$ in Eq.\,\eqref{eq:Delta1} is sloppy it is safe to conclude that the anomaly in the fluctuation disordered regime has a correspondence in the macroscopic properties while that at \Tc is of microscopic origin. In this way the temperature dependence corroborates the results with applied field indicating that the phonons alone sense the formation of helimagnetic order whereas the thermodynamic properties and the phonons are similarly affected by the fluctuations. The discrepancies between microscopic and macroscopic properties are largest at \Tc. However, they are significant also below. We recall that the  Gr\"uneisen parameter turns negative right above \Tc meaning that the phonon frequencies anomalously increase along with the volume. In the case of MnSi this anomaly in $\gamma_{{\rm macro}}$ can be traced back to the thermal expansion \cite{Matsunaga:1982,Pfleiderer:2007a}. The strong discrepancies between microscopic and macroscopic properties  highlight that the global volume changes observed around \Tc are insufficient to explain the phonon anomalies. Rather, there are interactions that leave a much stronger imprint on the phonons than on the overall bulk properties. %
In fact, Fawcett et al.~\cite{Fawcett:1970,Fawcett:1989} found large magnetic contributions in studies of the specific heat and the elastic properties below \Tc which they interpreted in terms of a magnetic Gr\"uneisen parameter $\gamma_{\rm mag}$. The authors found a $\gamma_{\rm mag}$ as large as -45 and constant in a temperature range from 14 to 32\,K except for significant deviations very near to \Tc. Also Pfleiderer and coworkers argue that there is a sizable anomalous contribution to the thermal expansion beyond the conventional $T^2$ variation\cite{Pfleiderer:2007a}.

\subsection{Carrier properties}
\label{sec:electronicRaman}

\begin{figure}[t]
\includegraphics[width=0.43\textwidth]{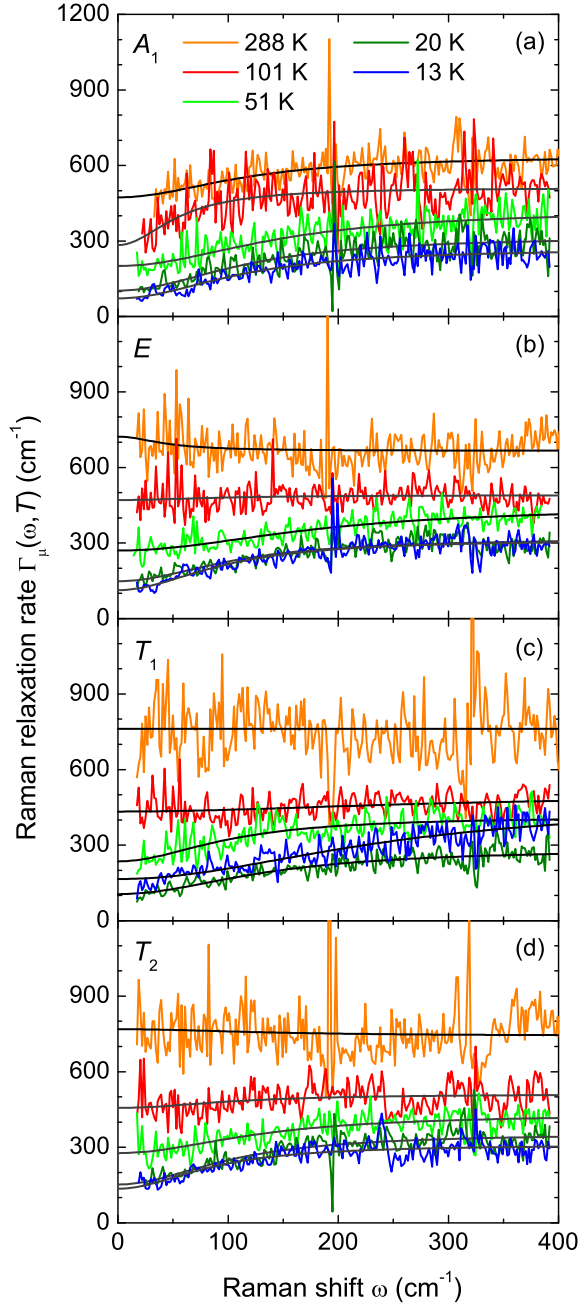}%
\caption{Temperature and symmetry dependence of the relaxation rates $\Gamma_{\mu}(\omega,T)$. The rates were obtained from the electronic continua as shown in Fig.~\ref{fig:continua}. To calculate $\Gamma_\mu(\omega,T)$ the procedure described by Opel and coworkers\cite{Opel:2000} was used. The smooth lines are phenomenological fits to the data according to Eq.~\eqref{eq:Gamma_old}.
}
\label{fig:Gamma}
\end{figure}

It is a standard procedure in optical (IR) spectroscopy to derive transport lifetimes $\tau (\omega,T)$ or scattering rates $\Gamma = 1/ \tau$ as well as the optical mass $m^{\ast}/m_b = 1+\lambda(\omega,T)$ from the reflectivity~\cite{Allen:1977,Tanner:1992} or from ellipsometry data. If the single particle relaxation depends on the momentum {\bf k}, as expected for MnSi~\cite{Belitz:2006}, IR yields only averages over the entire Fermi surface since the current response is typically not very {\bf k} selective, and some momentum resolution may be similarly useful as in the cuprates, for instance.\cite{Opel:2000}

\begin{figure}[t]
\includegraphics[width=0.43\textwidth]{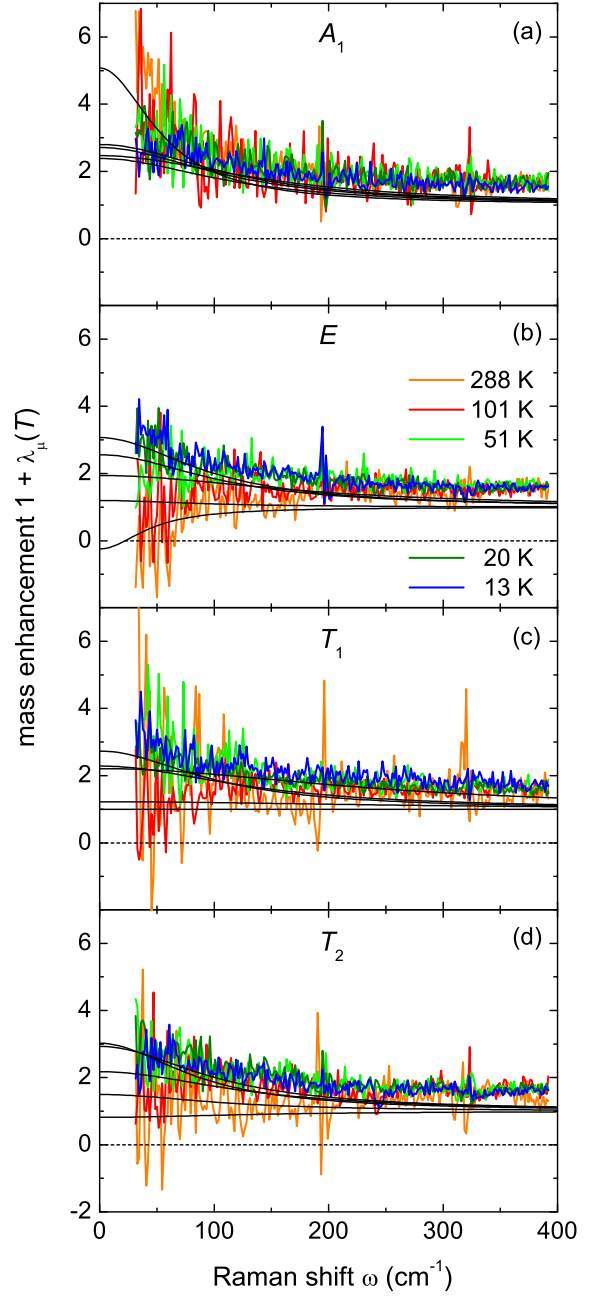}%
\caption{Temperature and symmetry dependence of the optical masses $1 + \lambda_{\mu}(T)$ obtained via the same formalism as the relaxation rates. The smooth lines are derived from the fits to the relaxation rates via Kramers-Kronig transformation, hence obey causality (see text). They are in reasonable agreement to the data except for a constant offset.
}
\label{fig:lambda}
\end{figure}

For this reason electronic Raman scattering can provide additional information.\cite{Devereaux:1994} For the normal state, Opel and coworkers demonstrated that the relaxation or memory function approach proposed by G\"otze and W\"olfle~\cite{Goetze:1972} can be adapted to facilitate the derivation of lifetimes $\tau_{\gamma\gamma}(\omega,T)$ or scattering rates $\Gamma_{\gamma\gamma} = 1 / \tau_{\gamma\gamma}$ from the electronic continuum as a part of the response  $R\chi_{\gamma\gamma}^{\prime \prime}$.~\cite{Opel:2000} Here, $\gamma$ is short hand for $\gamma({\bf k})$ and represents the Raman vertex or form factor which projects out symmetry dependent parts of the Brillouin zone. For simplicity and in order to avoid confusion with the Gr\"uneisen parameters, we label the derived quantities by the symmetry projection $\mu$ rather than the vertex $\gamma$.

Fig.~\ref{fig:Gamma} shows the dynamical relaxation rates $\Gamma_{\mu}(\omega,T)$ derived from the energy dependent response $R\chi^{\prime\prime}_{\mu}(\omega,T)$ (Fig.~\ref{fig:continua}) as described in Ref.~\onlinecite{Opel:2000}. They all have a similar temperature dependence at high energies but exhibit substantial differences close to zero energy. While $\Gamma_{T1}$, and $\Gamma_{T2}$ become rather flat at room temperature, $\Gamma_{A1}$ dips down at low energy thus reducing the overall temperature dependence between 13 and 288\,K. In contrast, $\Gamma_{E}(288\,\kelvin)$ increases slightly towards low energy. The rates reflect the variation of the raw data (Fig.~\ref{fig:continua}) but, owing to the derivation procedure\cite{Opel:2000}, show some features in a more pronounced fashion such as the low-energy variation with temperature.

For being derived from $R\chi^{\prime\prime}_{\mu}(\omega,T)$ in the same way, the dynamical mass renormalization factors $1+\lambda_{\mu}(\omega,T)=m_{\mu}^{\ast}(\omega,T)/m_b$, where $m_b$ is the band mass, are expected to similarly emphasize the variations close to zero energy. In fact, Fig.~\ref{fig:lambda}\,(b) shows an anomaly of $1+\lambda_{E}(\omega,T)$ at 288\,K. While all other masses increase monotonically with decreasing energy (other panels of Fig.~\ref{fig:lambda}), as typically expected for metals, $1+\lambda_{E}(\omega,288\,\kelvin)$ is reduced and even becomes negative at low energies. On the high energy side the masses saturate between 1 and 2 as expected. For the $A_1$, $T_1$, and $T_2$ symmetries the masses decay monotonically. In the zero-energy limit they reach temperature dependent values between 0.8 ($T_2$) and 5 ($A_1$).

The dynamical carrier properties found here with light scattering are in overall agreement with those derived from the reflectivity\cite{Mena:2003}. While the magnitudes and temperature dependences of $\Gamma_{\mu}(\omega,T)$ and $1+\lambda_{\mu}(\omega,T)$ are similar below 300\,K there are important differences which may lead to new insights: (i) The masses found here are generally above unity indicating the existence of interactions at all temperatures and energies. (ii) The symmetry dependence is significant. This is a unique feature of light scattering and indicates the existence of anisotropies in the Brillouin zone which cannot be derived from the optical conductivity.

However, as opposed to the cuprates\cite{Devereaux:1994,Devereaux:2007} or the iron-based compounds\cite{Muschler:2009,Mazin:2010a}, it is not possible to directly map the symmetries on separate bands or regions in the Brillouin zone since only three of the seven Fermi surfaces are in high symmetry positions around the $\Gamma$ $(0,0,0)$  and $R$ $(\pi,\pi,\pi)$ points\cite{Jeong:2004}. In particular, the minority and majority spin Fermi surfaces cannot be distinguished. According to the lowest order Brillouin zone harmonics for cubic crystals\cite{Allen:1976}, carriers on the Fermi surfaces around the $\Gamma$- and $R$-point may be projected out predominantly in $A_1$ symmetry. The four other Fermi surfaces have similar shapes and can be thought of as being built of three tubes intersecting around the $\Gamma$ point. This means that they should contribute to the spectra of all symmetries although different parts will be emphasized by different symmetries. For example, the $E$ spectra are dominated by the necks around the $X$ points $(\pi,0,0)$. More detailed information about the projections may be obtained by calculating the band curvatures which yield the sensitivities for a given band structure in the non-resonant case\cite{Devereaux:2007,Mazin:2010a}.

While the results in the $A_1$ and $T_2$ symmetries are compatible with metallic behavior at all temperatures the $E$ spectra deviate remarkably from what one expects for a metal. The deviation is seen best in the dynamics of the mass at 288\,K which varies non-monotonically. At room temperature this observation is paralleled by the IR results reported by Mena et al.~\cite{Mena:2003}. Not surprisingly, it survives down to lower temperatures in the Raman spectra since the Brillouin zone projections are more selective.

As to the interpretation, the relaxation rates are more intuitive. The increase towards zero energy indicates either a new relaxation channel or pseudogap-like behavior at higher temperatures. Similar anomalies have in fact been observed in organic conductors\cite{Toyota:2007} and cuprates\cite{Venturini:2002c}. For heavy Fermion systems this type of temperature dependence can be explained in terms of a Kondo-like interaction\cite{Chattopadhyay:1997}. However, it is unusual that the anomaly appears here at high temperature and vanishes below 100\,K. Obviously, at least parts of the Fermi surface exhibit insulating behavior at higher temperature. Whether or not this can be observed in ordinary transport remains open at the moment since there is no data available at elevated temperature. In addition, the metallic parts could short circuit the insulating ones in a similar fashion as in the cuprates\cite{Venturini:2002b}. We note that a vanishing or even negative mass [Fig.~\ref{fig:lambda}\,(b)] can also result from multiband effects which, however, need to be studied numerically on the basis of a realistic band structure.

We now focus on the static (DC) limit. To reliably extract the zero-energy extrapolation values of $\Gamma_{\mu}(\omega \rightarrow 0,T)$ and $1+\lambda_{\mu}(\omega \rightarrow 0,T)$ from the relatively noisy data (cf. section~\ref{sec:ExpMeth}) we use phenomenological functions having the correct analytical behavior in the limits $\omega \rightarrow 0$ and $\omega \rightarrow \infty$: (i) $\Gamma_{\mu}(\omega,T)$ (as opposed to the imaginary part of the single particle self energy $\Sigma^{\prime\prime}$) is a symmetric function, $\Gamma_{\mu}(-\omega,T) = \Gamma_{\mu}(\omega,T)$, (ii) $\lambda_{\mu}(\omega \rightarrow 0,T > 0)$ is finite, and (iii) $\Gamma_{\mu}(\omega,T)$ saturates at high energy. The latter condition is a restriction in the spirit of the Mott-Joffe-Regel limit which, strictly speaking, applies only for single-particle lifetimes. In the case of two-particle response functions there are contributions to the carrier response beyond the mean free path, and general statements as to the high-energy behavior become impossible.\cite{Hussey:2004} Since most of the relaxation rates derived here saturate, the introduction of a temperature dependent limiting value $\Gamma_{\mu}^{\max}(T)$ is justified experimentally but is not well supported theoretically.

On this basis, the minimal model is the parallel-resistor formalism with a quadratic energy dependence at $\omega \rightarrow 0$,\cite{Hussey:2006}
\begin{equation}
\label{eq:Gammainv}
\frac{1}{\Gamma_{\mu}(\omega,T)} = \frac{1}{\Gamma_{\mu}^{\ast}(\omega,T)} +\frac{1}{\Gamma_{\mu}^{\rm{max}}(T)}.
\end{equation}
where $\Gamma_{\mu}^{\ast}(\omega,T)= c(T) + a(T) \omega^2$ dominates at low frequencies while $\Gamma_{\mu}^{\rm{max}}(T)$ describes the high energy part. Inversion yields
\begin{equation}
\label{eq:Gamma_old}
\Gamma_{\mu}(\omega,T) = \frac{\left( c(T) + a(T) \omega^2 \right) \cdot \Gamma_{\mu}^{\rm{max}}(T)}{c(T) + a(T) \omega^2 + \Gamma_{\mu}^{\rm{max}}(T)},
\end{equation}
with the zero frequency limit $\Gamma_{\mu}(0,T)$ given by
\begin{equation}
\label{eq:Gamma0}
\Gamma_{\mu}(0,T) = \frac{c(T) \Gamma_{\mu}^{\rm{max}}(T)}{c(T) + \Gamma_{\mu}^{\rm{max}}(T)}.
\end{equation}
\begin{figure}
\includegraphics[width=0.44\textwidth]{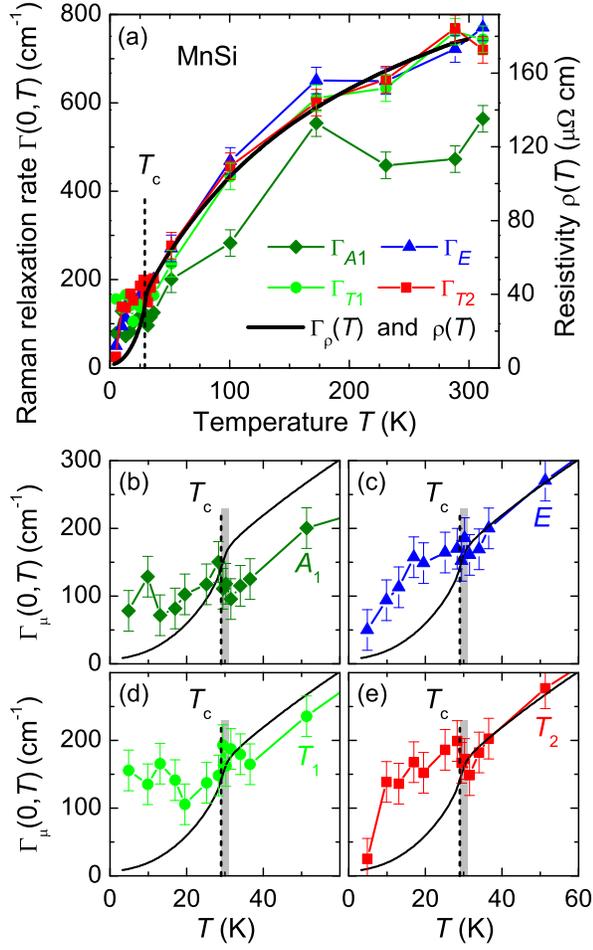}%
\caption{Static Raman relaxation rates and transport data. Panel (a) shows the Raman relaxation rates $\Gamma_{\mu}(\omega=0,T)$ (points, left axis) as a function of symmetry $\mu$ as derived from $\Gamma_{\mu}(\omega,T)$ (see text). If the longitudinal DC resistivity $\rho(T)$ (black line, right axis)\cite{Neubauer:2012} is converted into a relaxation rate $\Gamma_{\rho}(T)$ using a Drude model with the experimental plasma frequency $\omega_{\rm pl} = 2.3$\,eV~\cite{Mena:2003}, an extra factor of 0.73 is needed to match transport and Raman data. Above \Tc (dashed vertical line) the Raman data in $E$, $T_1$ and $T_2$ symmetry agree with $\rho(T)$, while the data in $A_1$ symmetry do not. (b)-(e) Zoom in on low temperatures. The phase transition has only a minor effect on $\Gamma_{\mu}(\omega=0,T)$. Above \Tc close to the fluctuation disordered regime (shaded), there may be a dip in $A_1$ and $T_2$ symmetry. Below the phase transition $\Gamma_{\mu}(\omega=0,T)$ decreases slower than $\Gamma_{\rho}(T)$.
}
\label{fig:transport}
\end{figure}

The fits to the relaxation rates $\Gamma_{\mu}(\omega,T)$ according to Eq.\,\eqref{eq:Gamma_old} are shown in Fig.\,\ref{fig:Gamma}. As expected, the $\omega \rightarrow 0$ extrapolation [Eq.\,\eqref{eq:Gamma0}] depends on both the high frequency limit $\Gamma_{\mu}^{\rm{max}}(T)$ and the offset $c(T)$. Each point in Fig.\,\ref{fig:transport} is obtained from such a DC extrapolation of $\Gamma_{\mu}(\omega,T)$.

The static Raman relaxation rates $\Gamma_{\mu}(0,T)$ can be compared to the longitudinal DC resistivity $\rho(T)$. To this end we show the resistivity measured on a comparable sample on the right axis of Fig.\,\ref{fig:transport}. In a Drude model the resistivity is related to the carrier relaxation rate $\Gamma_{\rho}(T)=\hbar/\tau(T)$ via the plasma frequency $\omega_{\rm pl}$ as $\Gamma_{\rho}(T) = \epsilon_0 \, \omega_{\rm pl}^2 \, \rho(T)$ with the vacuum permeability $\epsilon_0$ and $\omega_{\rm pl}=2.3$\,eV.~\cite{Mena:2003} An additional factor of 0.73 was introduced to match $\Gamma_{\rho}(T)$ and $\Gamma_{\mu}(\omega=0,T)$ (s.~Fig.\,\ref{fig:transport}).
A factor smaller than one can be explained by the frequency cutoff at 400\,cm$^{-1}$ in the relaxation rate analysis (Fig.\,\ref{fig:Gamma}). In addition, it cannot expected that the relaxation rates obtained from the light scattering experiment and from transport coincide completely since the higher order corrections to the respective response are different and in the ten percent range.\cite{Zawadowski:1990} The Raman data points $\Gamma_{\mu}(0,T)$ in $\mu=E$, $T_1$ and $T_2$ symmetry agree with $\Gamma_{\rho}(T)$ in the temperature range from 310 to 30\,K. The experimental error of about $\pm 30\,\centi\meter^{-1}$ was estimated from the scatter of neighboring points and the error of the DC extrapolation of $\Gamma_{\mu}(\omega,T)$, plotted in Fig.\,\ref{fig:Gamma}. It is remarkable that also $T_1$ symmetry, which essentially projects out chiral excitations, follows the resistivity curve at high temperatures. We argue that chiral excitations are allowed also above \Tc where no helimagnetic order is present, because also in the high temperature phase the crystal structure lacks inversion symmetry. Thus chiral excitations can be created with inelastic photon scattering even up to room temperature.

The $A_1$ symmetry component of the light scattering susceptibility, which has always an isotropic part, is screened by long range Coulomb interaction in the limit $\vq \rightarrow 0$. This is referred to as backflow, required to comply with particle number conservation and gauge invariance in crystals with free charge carriers~\cite{Devereaux:2007}. This screening effect may be one reason why the $A_1$ symmetry component of the DC relaxation rates does not match the resistivity data. Progress is probably only possible if the data are analyzed with a phenomenological model and on the basis of a realistic band structure.

Below \Tc the Raman relaxation rates for all symmetries stay well above the transport data and decrease at a lower rate but, finally become very small and, at least in $E$ and $T_2$ symmetry, approach values close to what one expects from the transport. The observation of this decrease makes it unlikely that the saturation observed in IR\,\cite{Mena:2003} or in the other symmetries is an artifact resulting from the rather small relaxation rates, and we conclude that the discrepancies in the ordered state need to be taken seriously. More specifically, the convex and the concave $T^2$-like behavior found in Raman and, respectively, in ordinary transport seem to be two sides of the same coin which need to be explained theoretically.

Similarly surprising is the anomaly right above \Tc which is particularly pronounced in $T_2$ symmetry [Fig.~\ref{fig:transport}\,(e)] but probably present also in $A_1$ and $E$ symmetry. It goes along with the dip in the phonon frequencies (Figs.~\ref{fig:phonon-E} and \ref{fig:phonon-T2}). The simultaneous observation of the anomaly in the phonon energy and in the electronic continuum at the same temperature makes us confident that the effect is significant. While the anomaly in the phonon channel is also observed independently in the Gr\"uneisen parameter as derived from the thermal expansion, the non-monotonic variation in the carrier relaxation is a new observation underpinning the impact of the phase transition and the preceding fluctuations on the electrons. The detailed analysis below 40\,K clearly reveals unexpected interactions between spin, charge, and lattice at the phase transition.

\section{Conclusions}
We have studied phonons and electronic excitations in MnSi by Raman spectroscopy. The phonons show conventional behavior above the helimagnetic transition but exhibit pronounced anomalies slightly above and below \Tc. Most remarkably, the phonon energy dips down already at 31\,K clearly above the transition. This anomaly is tracked by the Gr\"uneisen parameter. At \Tc the phonon energy first recovers and then softens before increasing at the lowest temperatures. Hence, the temperature dependence at and below \Tc has no correspondence in the thermodynamic properties and may indicate that the optical phonons scrutinized here couple to the spin order directly rather than via magnetostriction effects. In an applied magnetic field of 4\,T, in the field polarized state, the phonon anomalies disappear. This suggests a connection of the anomalies and helimagnetism.

The electronic relaxation rates in $E$ and $T_2$ symmetry agree reasonably well with conventional transport above 31\,K. Significant deviations are only found in the range $4<T<31$\,K. However theoretical support is required to understand these deviations from conventional transport and their symmetry dependence. The narrow minimum at $T_{\rm C}+2$\,K highlights the importance of the fluctuation range for the phase transition.

\begin{acknowledgments}
We acknowledge useful discussions with A. Baum, A.~F. Kemper, and B. Moritz and would like to thank A. Neubauer for the resistivity measurements. We are indebted to B.~S. Chandrasekhar for critically reading the manuscript and gratefully acknowledge the technical assistance of the WMI workshop. The work was supported by the DFG via the coordinated program PAK\,405 (project nos. HA\,2071/5-1 and PF\,393/10-1) and, partially, by the Transregional Collaborative Research Center TRR\,80.
\end{acknowledgments}

\bibliography{MnSi_paper,MnSi_Intro}

\pagebreak
\clearpage



\setcounter{figure}{0}
\makeatletter
\renewcommand{\thefigure}{A\@arabic\c@figure}
\makeatother

\begin{appendix}
\label{sec:appendix}

\section{Experimental setup}
\label{Asec:exp}

\begin{figure}[b]
\includegraphics[width=0.44\textwidth]{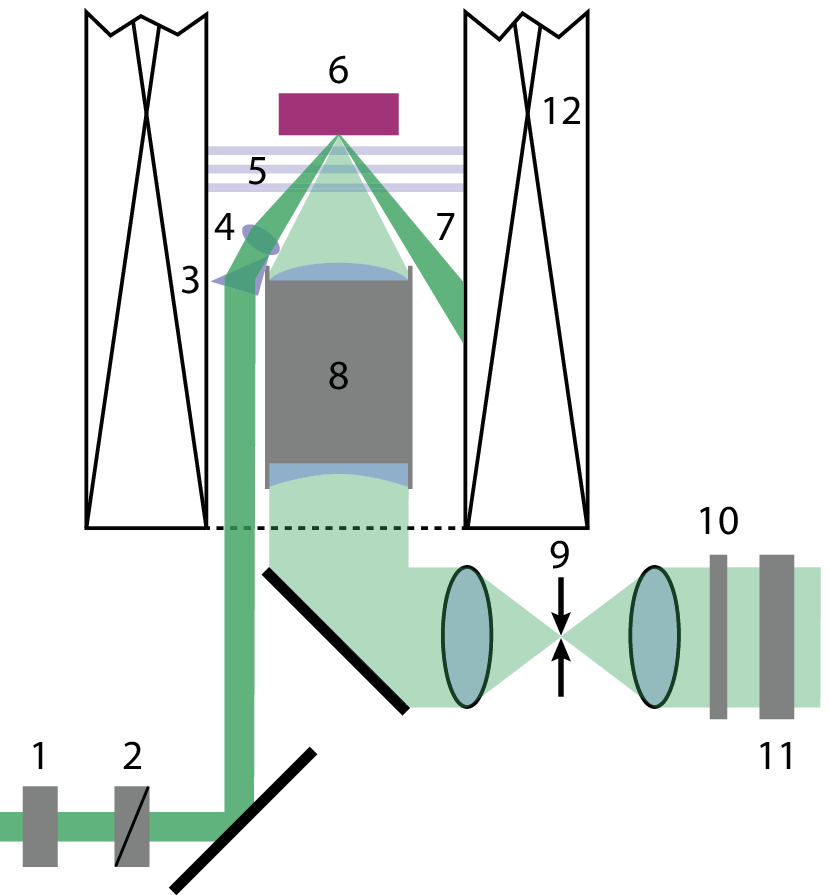}%
\caption{Sketch of the light path close to the sample. With a polarizer (1) and a Soleil-Babinet compensator (2) the polarization and phase of the incoming laser light (dark green) can be adjusted. The laser light is directed via a prism (3) to an achromat (4) focussing on the sample surface (6). Since separate optical components are used for excitation and collection, the direct reflex (7) from the mirror-like sample surface does not enter the collection optics (8) ($f \approx 30\,\milli\meter$, $N\!A = 0.34$) for the scattered light (light green). Reflexes from the cryostat windows (5) are blocked separately. This considerably reduces the amount of elastically scattered light and fluorescence reaching the spectrometer. The residual background signal is further reduced by a spatial filter (9) in the path of the scattered light consisting of two confocal achromatic lenses ($f = 75\,\milli\meter$) and a circular aperture of 150\,\micro\meter~in the common focus. A quarter wave plate (10) and an analyzer (11) select the polarization of the scattered light. The sample is located in a He-flow cryostat in the center of a superconducting solenoid (12).
}
\label{fig:light_path}
\end{figure}
For excitation a solid state laser emitting at 532.3\,nm was used. The light path in the vicinity of the sample is sketched in Fig.\,\ref{fig:light_path}. The incoming beam passes a polarizer (1) and a Soleil-Babinet compensator (2) to prepare the light polarization such that the photons inside the sample have the proper polarization state. Since the incoming light is focussed on the sample surface at an angle of $30\degree$, the absorption of photons polarized parallel and perpendicular to the plane of incidence is not equivalent. Circularly polarized light, for example, assumes an elliptical polarization inside the sample. The Soleil-Babinet compensator (2) takes care of this problem by facilitating independent access to polarization and phase~\cite{Prestel:2012}. In terms of the example above, the light is elliptically polarized outside the sample such that the absorbed light is circularly polarized inside the sample. To this end, one has to keep track of all optical elements in the light path. The major effect comes from the sample surface and is determined by the complex index of refraction $\hat{n}$ (cf. section~\ref{sec:samples}). Via a mirror and a prism (3) the photons are directed to an achromat (4) and focused on the sample surface (6). The angle of incidence of 30\degree~guarantees that the  photons reflected (7) off the sample surface do not enter the collection optics (8). In this configuration the contributions from fluorescence in the optics and from the laser light is minimized. The Raman light is collected by a custom-made objective lens (8) having a numerical aperture of $N\!A = 0.34$ corresponding to a solid angle of 0.37\,sr. For maximal throughput and best imaging properties the geometrical aberrations introduced by the cryostat windows (5) are corrected by the objective lens (8). In this way the scattered light can be spatially filtered (9) without excessive losses. The objective (8) and the spatial filter (9) are the key parts which facilitate measurements on samples with very few scattered photons. The scattering geometry is dictated by a solenoid (12) which allows us to apply magnetic fields up to 8\,T.

Before the scattered photons arrive at the entrance slit of the spectrometer the desired polarization states are selected with a quarter wave plate (10) and an analyzer (11). For dispersion a triple-stage spectrometer (Jobin-Yvon T64000) with the spectral resolution set at approximately 2\,\wn is used. The transmitted photons are recorded with a liquid nitrogen cooled charge coupled device (CCD) detector.

\section{Temperature determination}
\label{sec:T-determination}

\begin{figure}[b]
\includegraphics[width=0.44\textwidth]{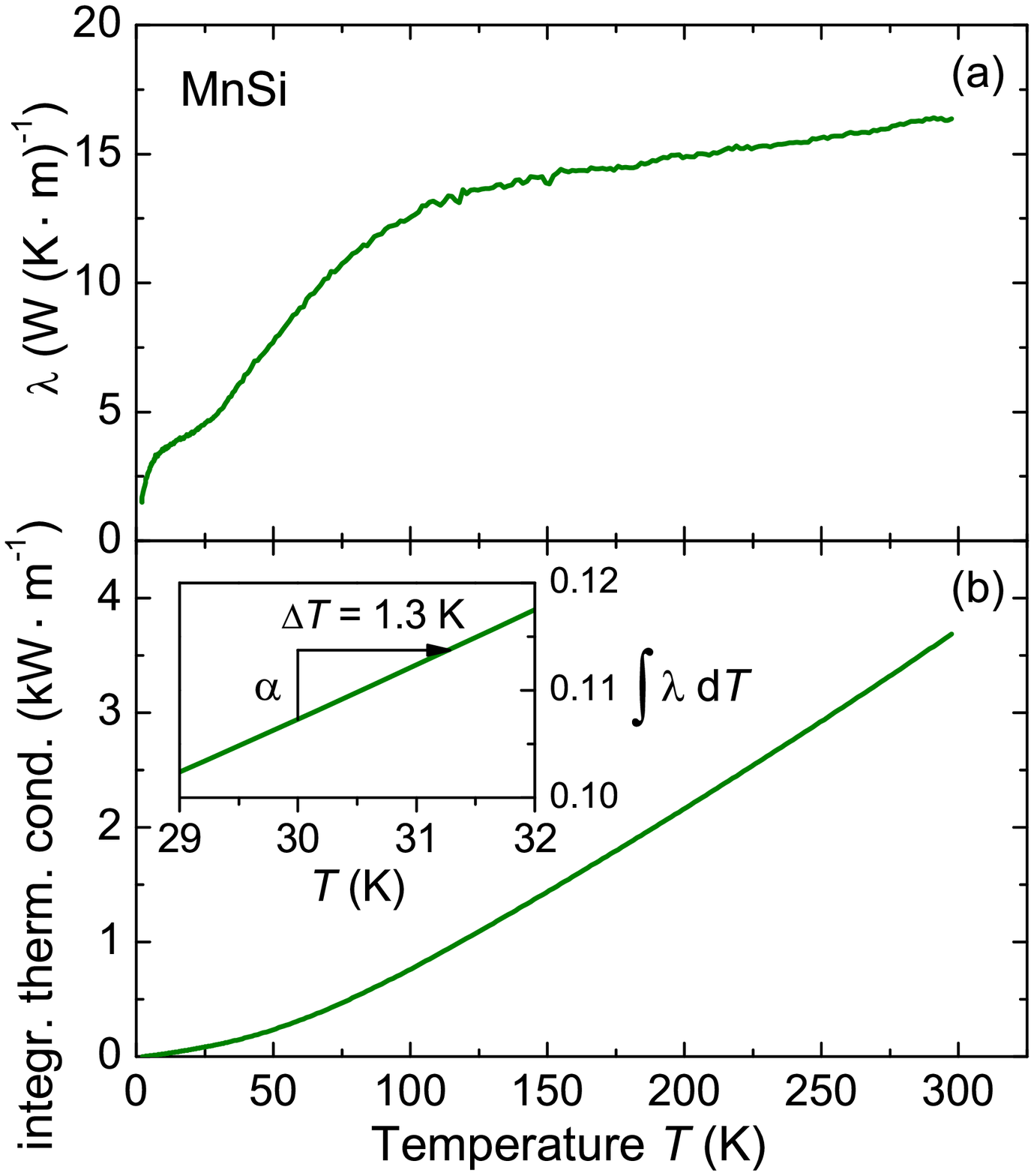}
\caption{Temperature determination from thermal conductivity. (a) thermal conductivity measurement; (b) integrated thermal conductivity. According to Eq.\,\eqref{eq:heating}, the integral $\int \lambda dT =: \alpha$ from $T_{\rm h}$ to $T_{\rm h} + \Delta T$ is constant as long as the laser power and the focus size are not changed. For the experimental setup used here, $\alpha$ is 6.4\,W/m for an absorbed laser power of 4\,mW. The inset of (b) illustrates how $\Delta T$ can be obtained from $\alpha$ and the integrated conductivity.
}
\label{fig:thermal_conductivity}
\end{figure}

The physical properties depend crucially on the temperature in the illuminated spot $T_{\rm{s}}$. The absorbed light heats the sample locally and results in a sizeable difference $\Delta T$ between the holder temperature $T_{\rm h}$ and $T_{\rm s}$, particularly at low temperatures. In the main article, $T_{\rm{s}}$ is referred to as $T$ for simplicity. A precise temperature determination is essential to decide if anomalies in the spectra close to \Tc are actually appearing above, at, or below the transition.

\begin{figure}
\includegraphics[width=0.44\textwidth]{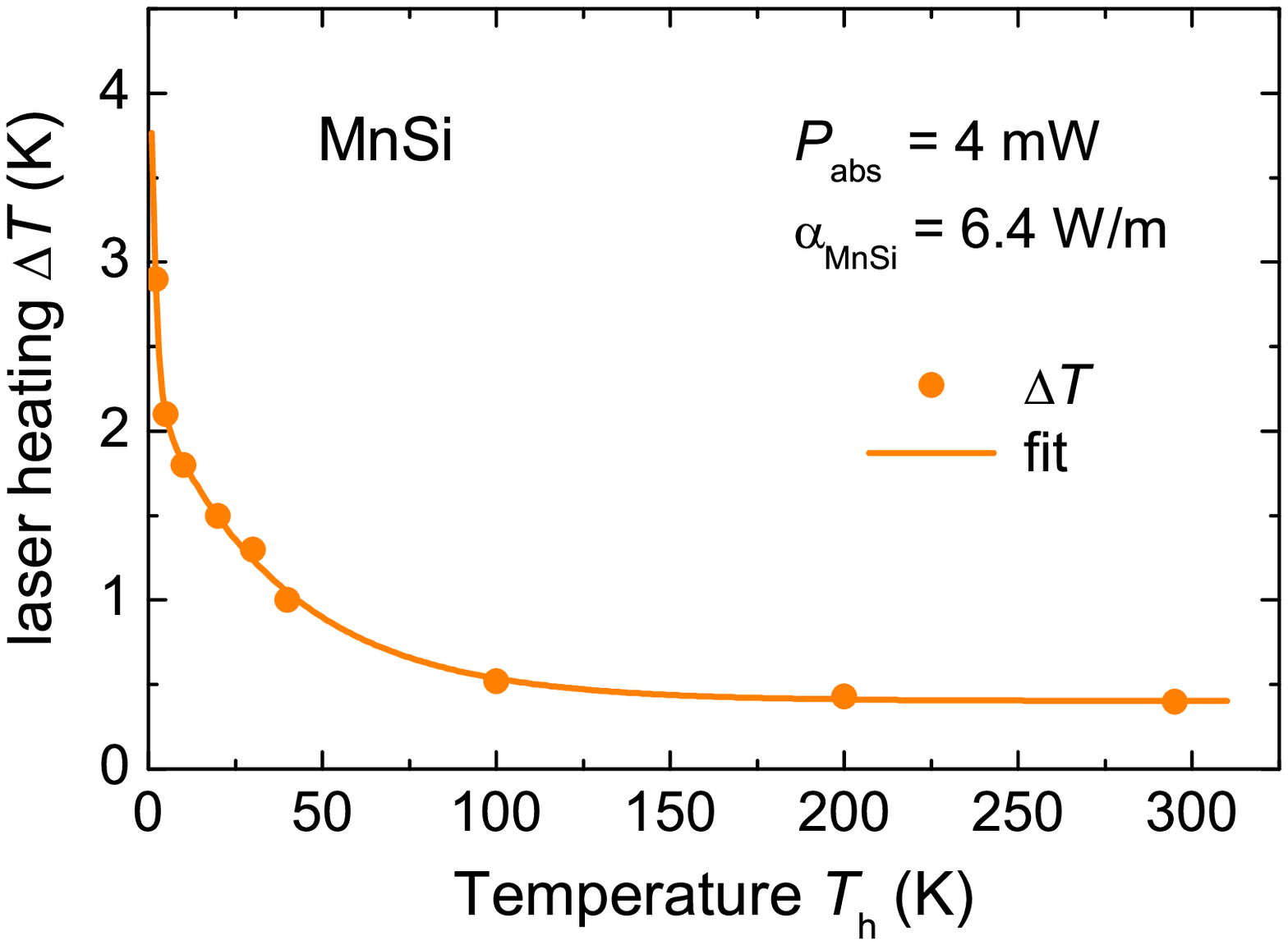}%
\caption{Temperature dependence of the laser heating $\Delta T$ for an absorbed laser power of 4\,mW. Points correspond to the laser heating for various holder temperatures \Th and are determined from the integrated thermal conductivity as described in Fig.\,\ref{fig:thermal_conductivity}. An exponential fit was used to obtain $\Delta T$ in the full temperature range.
}
\label{fig:laser_heating}
\end{figure}

One common method to determine the amount of heating is to compare Stokes and anti-Stokes intensity, i.e. energy gain and energy loss spectra. They differ by an exponential factor due to thermal occupation which can in principle be used to compute the temperature. However, this method is not always reliable, in particular at temperatures below 100\,K and in materials like MnSi with low scattering intensities. Alternatively, strongly temperature dependent features in the spectra can be analyzed using combinations of various laser powers $P_{\rm{abs}}$ and holder temperatures $T_{\rm h}$. Ideally, $P_{\rm{abs}}$ and $T_{\rm h}$ are selected in a way that the spectral features assume the same shape for at least two combinations. Then, if $\Delta T$ is moderate, one can expect that the spot temperatures are equal. However, there are no strongly temperature dependent features in the spectra of MnSi throughout the whole temperature range.

The laser heating used here was determined from previous results on V$_3$Si~\cite{Hackl:1983,Hackl:1987}. There, the strong temperature dependence of the gap mode in the superconducting state was used to determine the amount of laser heating. Together with data for the thermal conductivity $\lambda(T)$ the complete temperature range of a material becomes accessible. This method can be used for other materials having isotropic heat conduction by appropriately taking into account $\hat{n}$ and $\lambda(T)$. The extrapolation of this procedure worked for Nb$_3$Sn \cite{Hackl:1987}. Noticing that MnSi, while not having sufficiently temperature dependent features, is an isotropic metal, we apply that procedure and provide a few details.

Starting point is the equation for thermal conductivity in semi-spherical geometry. The laser power $P_{\rm abs}$ is deposited within the illuminated spot with radius $r_0$. It is assumed that the complete energy is transferred to the sample holder via heat conduction in the sample. This is an excellent approximation for metals in both vacuum and gas atmosphere, but certainly not when the sample is immersed in superfluid He. In the case of transport in the metal, the heat flow through a shell with area $2\pi r^2$ and thickness $dr$ is given by
\begin{equation}
P_{\rm{abs}} = - \lambda (T) \cdot 2 \pi r^2 \frac{dT}{dr},
\label{eq:thermCond}
\end{equation}
where $dT$ is the temperature drop across the radial increment $dr$. If the sample is large compared to $r_0$, integration yields\cite{Ready:1971,Hackl:1987}
\begin{equation}
\alpha(P_{\rm{abs}},r_0) := \frac{P_{\rm{abs}}}{2\pi \,r_0} = \int_{T_{\rm h}}^{T_{\rm h} + \Delta T} \lambda (T) dT.
\label{eq:heating}
\end{equation}
$T_{\rm{h}}$ is measured with a Cernox resistor having a vanishingly small magneto-resistance of less than 0.2\% for magnetic fields up to 4\,T. $\Delta T = T_{\rm{s}} - T_{\rm{h}}$ is the average laser-induced heating. $\alpha$ depends only on $r_0$ and $P_{\rm{abs}}$ but not on any strongly temperature dependent property of the sample. Hence the knowledge in one sample, such as V$_3$Si, is sufficient for deriving $\Delta T$ of other compounds. The main non-trivial sample dependence comes from $\lambda (T)$ which has to be known in detail. Then $\Delta T$ can be determined for any $T_{\rm{h}}$ if the integral over $\lambda (T)$ is known (cf. Fig.\,\ref{fig:thermal_conductivity}).

In V$_3$Si $\alpha_{\rm V3Si} = 0.96\,$W/m was derived experimentally for an absorbed laser power of 1\,mW. For MnSi we used $P_{\rm{abs}} = 4.0\,\milli\watt$ for all light polarizations. Differences in the optical setup of both experiments change the focus size by a factor of 0.6. Scaling $\alpha_{\rm V3Si}$ with these changed experimental parameters results in $\alpha_{\rm MnSi} = 6.4$\,W/m. As an example the graphical solution of Eq.\,\eqref{eq:heating} for the MnSi sample is shown in Fig.\,\ref{fig:thermal_conductivity}. The experimental quantity $\alpha_{\rm MnSi}$\,=\,6.4\,W/m for $P_{\rm abs}=4\,$mW is added to the integrated thermal conductivity at the holder temperature \Th. Then, a horizontal line intersects the integral $\int \lambda dT$ at the spot temperature \Ts. The length of this line is $\Delta T$. In the example of Fig.\,\ref{fig:thermal_conductivity}, \Th = 30\,K, \Ts = 31.3\,K and $\Delta T$ = 1.3\,K.

In the same way, the laser heating was determined at several different holder temperatures between 4 and 300\,K. These temperatures are indicated as points in Fig.\,\ref{fig:laser_heating}. An exponential fit yields $\Delta T$ for all temperatures in between and was used to correct for the laser heating in all measurements presented here. At selected holder temperatures $T_{\rm{h}} =\,$2, 30, and 300$\,\kelvin$, $\Delta T$ is 2.9, 1.3, and 0.4$\,\kelvin$, respectively, for $P_{\rm abs}=4\,\milli\watt$.

\section{Data analysis}
\label{Asec:analysis}

\begin{figure}[h]
\includegraphics[width=0.44\textwidth]{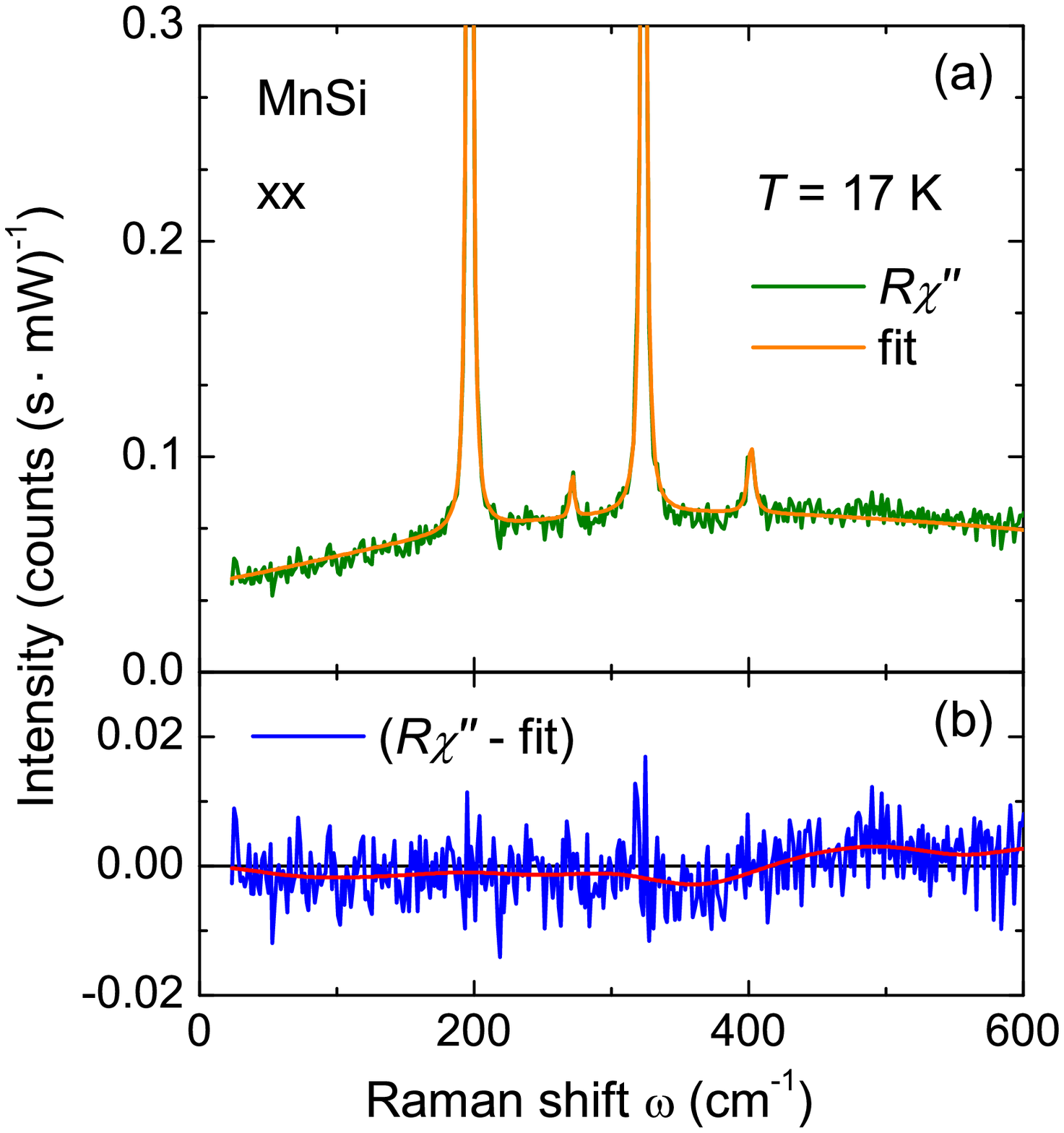}%
\caption{Fit of the data on the example of a $xx$ spectrum at 17\,K. Panel (a) shows the measured data (green) together with the fit function (orange) consisting of a 5th order polynomial and four Voigt shaped peaks. (b) shows the difference (blue) between measurement and fit. A smoothed curve (red) points out how much and at which frequencies the measurement and the fit differ beyond the noise level. To separate the phononic part from the electronic continuum, only the fits of the peaks are subtracted from the spectra.
}
\label{fig:voigt_fit}
\end{figure}

Another caveat in the analysis, connected with the low scattering intensity, is the fitting procedure used to separate the phononic contribution from the electronic continuum. Fig.\,\ref{fig:voigt_fit} shows a typical example. The Raman response $R\chi '' $ (green) is fitted (orange) by a polynomial baseline superimposed by several Voigt shaped peaks. The baseline is a 5th order polynomial fitted to 30 anchor points which are distributed in the full energy range, but not close to the peak positions. %
The baseline is subtracted to put the peaks on a horizontal ground for the Voigt fit. Panel (b) shows the difference (blue) between the data and the fit which is less than 0.005\,\cps for most points. The smoothed line (red) is a guide to the eye. To get the electronic continua shown in Fig.\,\ref{fig:continua}, only the Voigt part of the fit at zero background is subtracted.

\section{Gr\"uneisen parameter}
\label{Asec:Grueneisen}

\begin{figure}[b]
\includegraphics[width=0.44\textwidth]{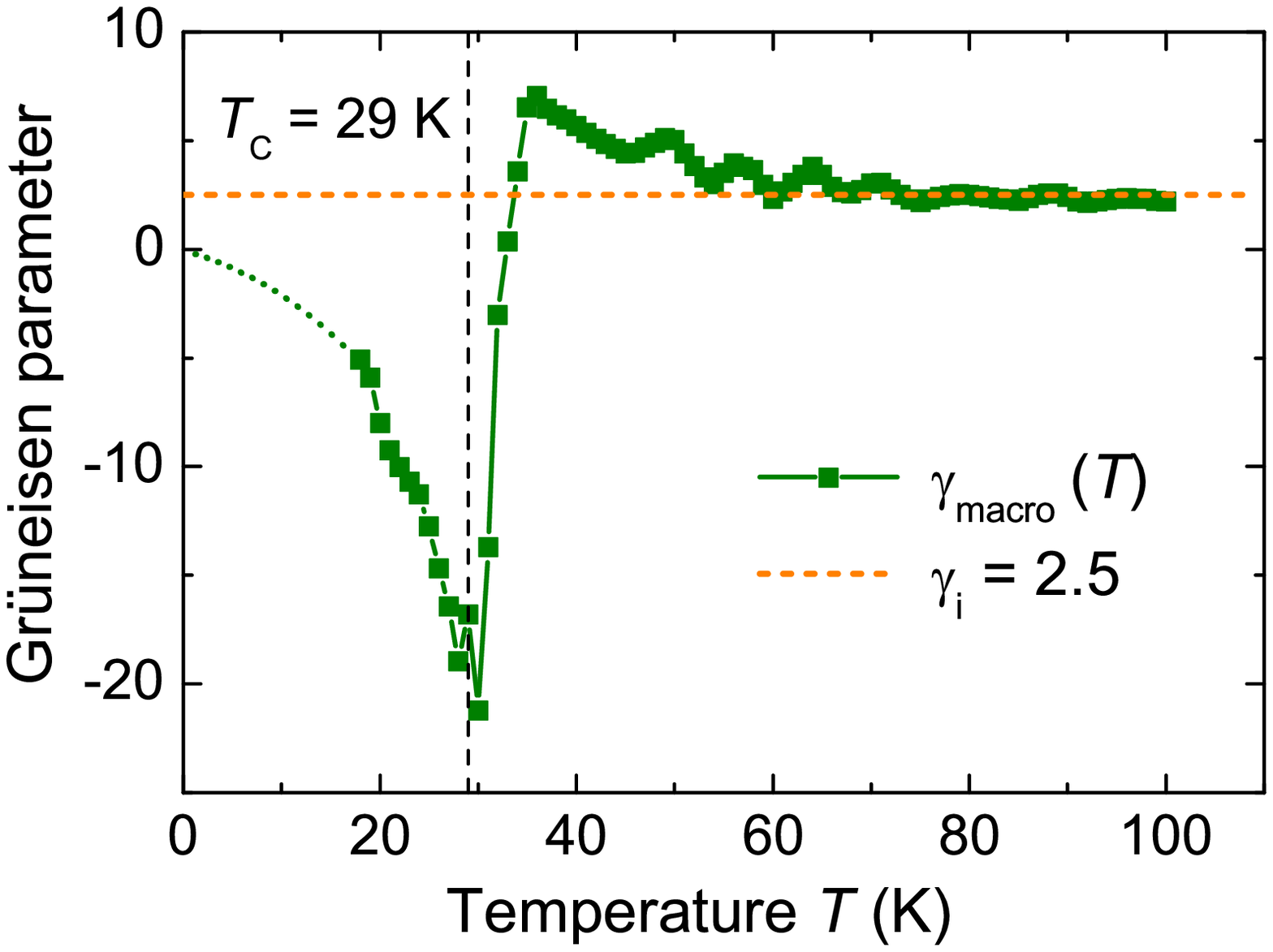}%
\caption{Gr\"uneisen parameters in MnSi. The macroscopic Gr\"uneisen parameter $\gamma_{\rm macro}(T)$ (green) as derived from thermodynamic properties in comparison with the constant phonon Gr\"uneisen parameter $\gamma_i=2.5$ (orange) which describes the thermal expansion shift of the Raman modes $i$.
}
\label{fig:Grueneisen}
\end{figure}

The macroscopic Gr\"uneisen parameter $\gamma_{\rm macro}$ can be determined from experimentally accessible thermodynamic properties\,\cite{AshcroftMermin:1988}.
\begin{equation}
\gamma_{{\rm macro}}(T)=\frac{3 \cdot \alpha(T) \cdot K(T) \cdot V^{\rm mol}(T)}{C_p^{\rm mol}(T)}
\end{equation}
The published data on thermal expansion, bulk modulus and heat capacity were sampled in 1\,K steps and then used to calculate $\gamma_{\rm macro}$\,\cite{Stishov:2008,Petrova:2009,Pauling:1948,2013:Bauer:PhysRevLett}. For $T \rightarrow 0$, $\gamma_{\rm macro}$ is expected to vanish (dots in Fig.\,\ref{fig:Grueneisen}) as $\alpha$ goes to zero. At temperatures below \Tc, $\gamma_{\rm macro}$ is negative with the minimum of about -20 at \Tc. Above the transition there is a steep increase and a sign change. $\gamma_{\rm macro}$ reaches a maximum of 7 at 36\,K and then approaches a value of 2.5 with increasing temperatures. This is compatible with the phonon Gr\"uneisen parameter $\gamma_i$ derived from the thermal expansion shift of the Raman modes at elevated temperatures (section \ref{sec:phonons}).

\end{appendix}

\end{document}